\documentclass[10pt]{article}
\usepackage{fullpage,graphicx,xspace,amsmath,amssymb,float}
\usepackage{lineno,cite,hyperref,authblk}
\usepackage[amssymb]{SIunits}
\usepackage{comment,multirow}

\hypersetup{
    colorlinks=true,  
    linkcolor=blue,   
    citecolor=blue,   
    filecolor=blue,   
    urlcolor=blue     
}

\newcommand{\be}{\[}
\newcommand{\ee}{\]}
\newcommand{\ben}{\begin{equation}}
\newcommand{\een}{\end{equation}}

\newcommand{\me}{\mathrm{e}}

\addunit{\hpi}{hpi}
\addunit{\ffu}{FFU}
\addunit{\hcvr}{HCV\ RNA}
\addunit{\Vu}{\ffu/\milli\liter}
\addunit{\Vscale}{V_{scale}}
\addunit{\Ru}{\hcvr/\micro\gram\ total\ RNA}
\addunit{\ssrpt}{SSR/pt}
\addunit{\perh}{\hour^{-1}}

\newcommand{\timpe}{Timpe et al.\ \cite{timpe08}\xspace}
\newcommand{\sainz}{Sainz et al.\ \cite{sainz06}\xspace}
\newcommand{\keum}{Keum et al.\ \cite{keum12}\xspace}
\newcommand{\graw}{Graw et al.\ \cite{graw15}\xspace}

\newcommand{\fI}{\ensuremath{f_{I(\unit{72}{\hour})}}\xspace}
\newcommand{\ssr}{\ensuremath{\mathrm{SSR}}\xspace}
\newcommand{\frinse}{\ensuremath{f_\text{rinse}}\xspace}
\newcommand{\Vo}{\ensuremath{V(0)}\xspace}
\newcommand{\Vopre}{\ensuremath{V(\unit{1}{\hpi})_\text{pre-rinse}}\xspace}
\newcommand{\Vopost}{\ensuremath{V(\unit{1}{\hpi})_\text{post-rinse}}\xspace}
\newcommand{\tinfect}{\ensuremath{t_{j,\text{infected}}}\xspace}
\newcommand{\trnaoffset}{\ensuremath{t_\text{RNAoffset}}\xspace}
\newcommand{\Pinf}{\ensuremath{\mathcal{P}_\text{infected}}}
\newcommand{\Pnifv}{\ensuremath{\mathcal{P}_\text{not-infected}^\text{free-virus}}}
\newcommand{\Pnicc}{\ensuremath{\mathcal{P}_\text{not-infected}^\text{cell-to-cell}}}
\newcommand{\Pifv}{\ensuremath{\mathcal{P}_\text{infected}^\text{free-virus}}}
\newcommand{\Picc}{\ensuremath{\mathcal{P}_\text{infected}^\text{cell-to-cell}}}
\newcommand{\pvec}{\ensuremath{\vec{p}} = (\trnaoffset, \ensuremath{c}, \ensuremath{\beta_V}, \ensuremath{\beta_C}, \Vo, \frinse, RNS)\xspace}

\title{Quantifying the relative contribution of\\ free virus and cell-to-cell transmission routes\\ to the propagation of hepatitis C virus infections in vitro\\ using an agent-based model}

\author[1]{Kenneth Blahut}
\author[1]{Christian Quirouette}
\author[2]{Jordan J.\ Feld}
\author[3,4,5,6]{Shingo Iwami}
\author[1,7]{Catherine A.A.\ Beauchemin\footnote{Corresponding author. Email: \url{cbeau@ryerson.ca}.}}

\affil[1]{Department of Physics, Ryerson University, Toronto, ON, Canada}
\affil[2]{Toronto Centre for Liver Disease, University Health Network, Sandra Rotman Centre for Global Health, University of Toronto, Toronto, Canada}
\affil[3]{Department of Biology, Kyushu University, Fukuoka, Japan}
\affil[4]{Institute for the Advanced Study of Human Biology (ASHBi), Kyoto University, Kyoto, Japan}
\affil[5]{NEXT-Ganken Program, Japanese Foundation for Cancer Research (JFCR), Tokyo, Japan}
\affil[6]{Science Groove Inc., Fukuoka, Japan}
\affil[7]{Interdisciplinary Theoretical and Mathematical Sciences (iTHEMS), RIKEN, Wako, Japan}

\date{\today}

\begin{document} 

\maketitle

\begin{abstract}
Experiments have shown that hepatitis C virus (HCV) infections in vitro disseminate both distally via the release and diffusion of cell-free virus through the medium, and locally via direct, cell-to-cell transmission. To determine the relative contribution of each mode of infection to HCV dissemination, we developed an agent-based model (ABM) that explicitly incorporates both distal and local modes of infection. The ABM tracks the concentration of extracellular infectious virus in the supernatant and the number of intracellular HCV RNA segments within each infected cell over the course of simulated in vitro HCV infections. Experimental data for in vitro HCV infections conducted in the presence and absence of free-virus neutralizing antibodies was used to validate the ABM and constrain the value of its parameters. We found that direct, cell-to-cell infection accounts for 99\% (84\%--100\%, 95\% credible interval) of infection events, making it the dominant mode of HCV dissemination in vitro. Yet, when infection via the free-virus route is blocked, a 57\% reduction in the number of infection events at \unit{72}{\hpi} is observed experimentally; a result consistent with that found by our ABM. Taken together, these findings suggest that while HCV spread via cell-free virus contributes little to the total number of infection events in vitro, it plays a critical role in enhancing cell-to-cell HCV dissemination by providing access to distant, uninfected areas, away from the already established large infection foci.
\end{abstract}


\cleardoublepage
\section{Introduction}

Virus infections in vitro can disseminate distally via release and diffusion of cell-free virus through the infection medium, and/or locally by direct cell-to-cell transmission of virus \cite{mothes10}. Each mode of infection presents various evolutionary advantages and disadvantages to the virus in vivo. Cell-free virions released into the extracellular medium can diffuse over large distances to infect cells, but are exposed and vulnerable to the innate and humoral immune responses as well as to antivirals targeting cell-free virus infectivity or the free virions' mode of attachment and entry. In contrast, cell-to-cell infection proceeds through close contact, with virions being transmitted directly from an infected cell to an adjacent uninfected cell. This direct transmission can protect virions from extracellular harm, but it also limits the pool of target cells available to the virions, restricting the spatial extent of infection. Many viruses, including the human immunodeficiency virus type 1 \cite{gupta89}, the murine leukemia virus \cite{jin09}, the respiratory syncytial virus (RSV) \cite{huong16}, and the hepatitis C virus (HCV) \cite{timpe08}, take advantage of both modes of infection.

Given the distinct manner by which cell-free and cell-to-cell HCV dissemination can affect and modulate the efficacy of a therapeutic antiviral intervention or that of the host immune response, it is important to resolve the relative contribution of either mode to the overall infection course. Unfortunately, this is particularly challenging to do in vivo, and is nearly as challenging to resolve in vitro for a number of reasons. Direct cell-to-cell HCV dissemination, at least in vitro, is thought to proceed via many of the same cellular receptors used for cell-free transmission including SR-BI, CLDN1, OCLN and NPC1L1 \cite{timpe08, brimacombe11, barretto14, sainz12nc1}, with some debate on the necessity of CD81 \cite{witteveldt09}. However, the route of direct, cell-to-cell transmission (e.g.\ gap junction, tight junctions, apical surface) and the nature of the matter being transmitted cell-to-cell (e.g.\ viral RNA segments, fully-formed encapsidated virion, host-enveloped lipoviral particles) have yet to be identified \cite{catanese13pnas}. As such, without knowing the pathways and infectious units specific to each mode of transmission, it is not possible to develop probes or markers to tag infection events and identify which mode caused each infection event. High-density, low-diffusion agar medium or neutralizing antibodies (nAbs), which significantly limit or inhibit cell-free virus dissemination, have been used to perform in vitro HCV infection experiments and evaluate the role of cell-free infection relative to that of direct cell-to-cell infection \cite{timpe08, graw15, barretto14, catanese13jvi}. Co-cultures have also been used to control or restrict HCV dissemination to the free-virus or cell-to-cell route individually \cite{witteveldt09, meredith13, brimacombe11, xiao14}. However, given that blocking cell-free infection modulates (hinders or facilitates) cell-to-cell transmission, and vice-versa, an experimentally-observed reduction in the number of infection events when one mode of transmission is blocked cannot simply, directly be ascribed to the contribution of that mode to infection.

In the case of HIV infections, these experimental limitations were overcome by using a mathematical model to analyze and appropriately interpret the experimental results \cite{iwami15}. Through mathematical analysis of experimental infections, Iwami et al.\ established that direct, cell-to-cell transmission of HIV was found to account for $\sim$60\% of all infection events in the dissemination of HIV infections in vitro. This was a particularly significant finding in light of an earlier experimental observation that the direct, cell-to-cell route of HIV transmission is resistant to some HIV antiviral therapy \cite{sigal11}. In the case of HCV, the relative contribution of the cell-free versus cell-to-cell modes of transmission has yet to be clearly resolved. Monotherapy with direct-acting antivirals (DAAs), which have replaced the previous standard-of-care treatment consisting of interferon-$\alpha$ and ribavirin therapy, are known to result in the emergence of DAA-resistant HCV mutants in vivo \cite{sarrazin10}. Importantly, these resistant virus appear to predominantly disseminate via the cell-to-cell route in vitro \cite{xiao14}. These recent observations highlight the importance of resolving the relative contribution of the cell-to-cell and cell-free routes of HCV transmission. In this work, we make use of an agent-based model (ABM) to simulate the course of in vitro HCV infections. The ABM is calibrated so as to reproduce the course and outcome of experimental HCV infections performed in the presence and absence of nAbs. The calibrated ABM is then used to identify the relative contributions of the two routes of HCV transmission in vitro.

 
\cleardoublepage
\section{Results}

\subsection{An agent-based model of HCV spread in vitro}

\begin{figure}
\begin{center}
\resizebox{\textwidth}{!}{\includegraphics{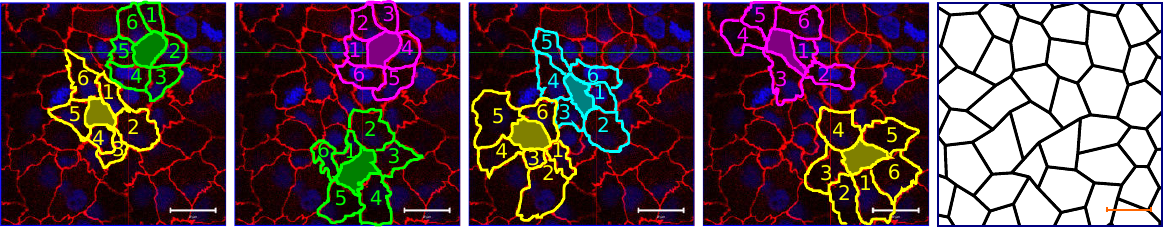}}
\caption{%
\textbf{Hexagonal packing of Huh7 hepatocytes in 2-D cell monolayers in vitro.}
The same image is repeated $4\times$, and in each view two central (shaded) cells are identified in one specific colour, and the cell's neighbourhood, i.e.\ those cells which appear to be in direct contact with that central cell, are coloured identically and numbered. Most cells appear to be in direct contact with 6 other cells, hence the choice of a hexagonal grid for the ABM, shown in the right-most panel (screenshot from the ABM rendered grid). The cell-culture image was captured by confocal microscopy (630$\times$, bar = \unit{20}{\micro\metre}) where cell nuclei (blue) were stained by Hoechst dye, and antibodies specific for tight junction protein zona occludens 1 (ZO-1, red) appear cleanly localized to intercellular junctions (see Methods for experimental details). The image has been adapted, as described in Methods, from the original article \textit{``Three-dimensional Huh7 cell culture system for the study of Hepatitis C virus infection''} by [Bruno Sainz Jr., Veronica TenCate, Susan L.\ Uprichard]. Virology Journal 2009, 6:103 (doi:\url{10.1186/1743-422X-6-103}) \cite{sainz09vj}. The original article is an open access article distributed under the terms of the Creative Commons Attribution License (\url{http://creativecommons.org/licenses/by/2.0}), which permits unrestricted use, distribution, and reproduction in any medium, provided the original work is properly cited.
}
\label{fhexgrid}
\end{center}
\end{figure}

We constructed an ABM which reproduces the spread of HCV infections in vitro by capturing HCV dissemination both locally, via cell-to-cell transmission, and distally, via the release and diffusion of free virus through the extracellular medium. The standard experimental in vitro infection system, consisting in a monolayer of hepatocytes, is represented in the ABM as a hexagonal grid wherein each site, $j=1,2,...,N_\text{cells}$, corresponds to one hepatocyte. The ABM grid is ($100\times100$) hepatocytes corresponding to a real-life spatial extent of approximately \unit{(2\times2)}{\milli\metre}, assuming one hepatocyte is \unit{20}{\micro\metre} in diameter. The hexagonal grid geometry was chosen because it best represents the natural arrangement and contact neighbourhood of confluent hepatocytes in cell cultures in vitro, as shown in Figure \ref{fhexgrid}. Graphically, the ABM grid is first rendered as a regular hexagonal lattice, and then each node is displaced by a random, normally distributed distance with zero mean and a standard deviation equal to 1/3 the hexagons' regular edge length. The node displacements have no impact on the resulting infection kinetics because they do not alter which hepatocyte neighbour which, but yield a more visually realistic cell layout for display purposes.

The ABM tracks the number of intracellular HCV RNA copies contained within each infected hepatocyte over time, $R_j(t)$ in hepatocyte $j$ at time $t$, expressed in units of HCV RNA copies. In contrast, the concentration of extracellular (cell free) infectious HCV is assumed uniform everywhere across the simulation grid (or cell culture), and is represented in the ABM as a single, global variable, $V(t)$, expressed in units of focus forming units or \ffu/\milli\liter. We tested and determined (not shown) that the infection time course under this assumption of a spatially uniform virus concentration, $V(t)$, does not differ significantly from that obtained when tracking local virus concentration at each site, $V_j(t)$, as we did previously in \cite{beauchemin06icaris}. This is not surprising given that the diffusion rate of HCV (\unit{10^{-11}}{\metre^2/\second}, based on Stokes-Einstein for HCV virions \unit{60}{\nano\metre} in diameter, assuming a viscosity of \unit{7\times10^{-4}}{\kilo\gram/(\metre\cdot\second)} for water at 37\degree\Celsius) is sufficiently large for this assumption to hold, as shown in our previous work \cite{holder11autoimm}.

In this way, the ABM-predicted, cell-free, extracellular, infectious HCV concentration in the medium (\ffu/mL), and the average intracellular HCV RNA level per cell (HCV RNA per $\micro\gram$ of cellular RNA) over the course of our in silico HCV infections, can be directly compared against these same experimental quantities commonly measured in vitro. One important advantage of the in silico infection model, however, is its ability to track, label, and distinguish between hepatocytes infected via the free virus versus the cell-to-cell routes of transmission, something we exploit later.

Both experimentally and in our ABM, the HCV infection is initiated by incubating hepatocytes with an inoculating medium containing extracellular infectious virus (\ffu/mL). If the hepatocyte at site $j$ becomes infected at time $t=\tinfect$, the number of HCV RNA copies within the newly infected hepatocyte, $R_j(t)$, grows at rate $\alpha$, following a Gompertz curve, to ultimately reach a steady state value of $\bar{R}$ HCV RNA copies. This is expressed in the ABM as
\ben
\hspace{-1em} R_j(t) = \begin{cases}
\bar{R}^{1-\exp\{-\alpha \left[t-(\tinfect + \trnaoffset)\right]\}} & \text{if } t > \tinfect \\
0 &  \text{otherwise}
\end{cases}\ \ ,
\label{intraRNA}
\een
where the age of infection of hepatocyte $j$ is given by $a_j = t-\tinfect$. An additional time offset, \trnaoffset, was introduced to account for the delay between successful HCV infection and the start of significant HCV RNA replication, as observed in \keum. This delay in viral production, referred to as the eclipse phase \cite{beauchemin17,holder11delay,holder11autoimm}, is implemented in Eqn.\ \eqref{intraRNA} such that $R_j(t=\tinfect+\trnaoffset)=R_j(a_j=\trnaoffset)=\unit{1}{\hcvr}$.

As the number of HCV RNA within an infected hepatocyte increases towards $\bar{R}$, so do the probability of direct cell-to-cell infection of a neighbouring cell and the rate of release of newly produced, infectious HCV progeny into the extracellular medium. The rate of release of infectious HCV into the medium by all hepatocytes in the ABM is represented as
\be
p_V(t) = \frac{\bar{V}\cdot c}{\bar{R} \cdot N_\text{cells}} \cdot \sum_{j=1}^{N_\text{cells}} \begin{cases}
R_j(t) & \text{if } t > (\tinfect + \trnaoffset) \\
0 &  \text{otherwise}
\end{cases} \ ,
\ee
where $\sum_{j=1}^{N_\text{cells}} R_j(t)$ for $t>(\tinfect + \trnaoffset)$ is the sum of intracellular HCV RNA over all infected hepatocytes as per Eqn.\ \eqref{intraRNA}, $N_\text{cells}$ is the number of hepatocytes considered in the ABM, $\bar{V}$ is the steady-state, cell-free infectious HCV concentration in the extracellular medium, and $c$ is the rate at which infectious HCV loses infectivity. The maximum rate of release of free infectious HCV into the medium, $p_V^\text{max} = \bar{V}c$, is reached when all hepatocytes have been infected for sufficiently long that all have reached their steady state intracellular HCV RNA concentration of $R_j = \bar{R}$ for all $j$. The concentration of cell-free, infectious HCV in the medium over the course of the infection, as per \cite{beauchemin06icaris,holder11autoimm}, is given by
\ben
V(t+\Delta t) = V(t) \,\me^{-c\Delta t} + p_V(t) \Delta t \ ,
\label{virevol}
\een
where $\Delta t$ is the duration of one time step of our ABM simulation, which is adapted at each step, chosen so as to ensure the probability of infection is less than 10\% in any one ABM step. The term $\me^{-c\Delta t}$ accounts for the reduction in the concentration of infectious HCV RNA in the medium over the course of one ABM time step, i.e.\ from $V(t)$ to $V(t+\Delta t)$, due to the exponential loss of HCV infectivity at rate $c$. When the rate of infectious HCV release into the medium is maximal, $p_V^\text{max} = \bar{V}c$, it is equal to the rate of loss of virus infectivity, resulting in a steady state of extracellular infectious HCV concentration in the medium of $\bar{V}$.

After hepatocytes infected by the initial inoculum have progressed through the initial phases of virus replication and release, the following rounds of infection can proceed. In our ABM, the probability that a hepatocyte \emph{does not} become infected via the free virus route decreases, as the concentration of extracellular infectious HCV in the medium increases, namely
\ben
\Pnifv(t) = \exp\left[-\Delta t \beta_V \frac{V(t)}{\bar{V}} \right] \ ,
\label{pnifv}
\een
where the steady state concentration of extracellular infectious HCV, $\bar{V}$, is used to rescale $V(t)$ such that $\beta_V$, a rate representing extracellular HCV infectivity per unit time, is unaffected by and independent of the method used to measure it experimentally (e.g.\ \ffu\ vs TCID$_{50}$). Similarly, the probability that a hepatocyte \emph{does not} become infected via the cell-to-cell route decreases as the fraction of the cell's six neighbours which have been infected for longer than time $\trnaoffset$ increases, namely
\ben
\Pnicc(t) = \exp\left[-\Delta t \beta_C \frac{\sum_{\text{nei}=1}^6 I_\text{nei}(t)}{6} \right] \text{  where  }
I_\text{nei}(t) = \begin{cases}
1 &  t > (\tinfect + \trnaoffset) \\
0 & \text{otherwise}
\end{cases}\ ,
\label{pnicc}
\een
where $\beta_C$ is a rate representing cell-to-cell HCV infectivity per unit time, analogous to $\beta_V$ for cell-free infections. As such, the probability that a hepatocyte is infected by either mode of infection (free virus or cell-to-cell) is given by
\be
\Pinf(t) = 1 - \left[ \Pnifv(t) \cdot \Pnicc(t) \right] \ ,
\ee
and the probability that a hepatocyte is infected by one specific mode of infection is given for each mode by
\begin{align*}
\Pifv &= \Pinf \cdot \left( \frac{1-\Pnifv}{2-\Pnifv-\Pnicc} \right) \\
\Picc &= \Pinf \cdot \left( \frac{1-\Pnicc}{2-\Pnifv-\Pnicc} \right) \ .
\end{align*}
In all, our ABM has 3 biological parameters related to intracellular HCV replication ($\bar{R}$, $\alpha$, \trnaoffset) from Eqn.\ \eqref{intraRNA}, and 4 biological parameters ($\bar{V}$, $c$, $\beta_V$, $\beta_C$) related to the kinetics of HCV dissemination from Eqns.\ \eqref{virevol},\eqref{pnifv},\eqref{pnicc}. These parameters will need to be determined by challenging the ABM with experimental in vitro HCV infection data, which we tackle in the next section.

\subsection{Estimating ABM parameters from in vitro HCV infection data}

Three distinct experimental data sets were used to constrain the biologically-relevant parameters of the ABM, and to ultimately determine the relative contributions of the cell-free virus versus cell-to-cell modes of infection to HCV dissemination in vitro: \keum, \sainz, and \timpe. Table \ref{tdata} provides an overview of the source and nature of the data, and how they were used herein to parametrize the ABM.

\begin{table}
\caption{Summary of data used in the analysis.}
\newcommand{\txcolwi}{0.34\textwidth}
\newcommand{\useabm}{estimate ABM PPLDs via MCMC with Eqn \eqref{elikeli}}
\label{tdata}
\vspace{1em}
\begin{tabular}{p{0.223\textwidth}|p{0.181\textwidth}|p{0.183\textwidth}|p{\txcolwi}}
\multicolumn{1}{c}{Source of data} & \multicolumn{1}{c}{Figure herein} & \multicolumn{1}{c}{Experiment details} & \multicolumn{1}{c}{Application} \\
\hline
\hline
\multicolumn{4}{l}{Intracellular positive strand HCV RNA/cell}  \\
\hline
\keum & & MOI: 6 $\ffu$/cell & \multirow{3}{\txcolwi}{\raggedright Kinetics of HCV replication within a single cell. Fitted with Eqn.\ \eqref{intraRNA} to estimate and fix parameter $\alpha$.} \\
Fig.\ 1A (black diamond) & Fig.\ \ref{fkeum12} (red circles) & Virus: JFH-m4$^a$  & \multirow{3}{\txcolwi}{} \\
 &  & Cell line: Huh7.5.1  & \multirow{3}{\txcolwi}{} \\
\hline
\hline
\multicolumn{4}{l}{Intracellular HCV RNA/$\micro\gram$ total cell RNA } \\
\hline
\sainz & & MOI: 0.01 $\ffu$/cell & \multirow{4}{\txcolwi}{\raggedright Intracellular HCV kinetics during multiple-cycle infection. Used to fix $\bar{V} = \unit{10^4}{\frac{FFU}{mL}}$ and \useabm.} \\
Fig.\ 3A (open diamond) & Fig.\ \ref{freprod}A (red stars) & Virus: JFH1 & \multirow{4}{\txcolwi}{} \\
Fig.\ 4A (line) & Fig.\ \ref{freprod}A (red circles) & Cell line: Huh7  & \multirow{4}{\txcolwi}{} \\
 & & & \\
\hline
\multicolumn{4}{l}{Extracellular infectious HCV in $\ffu/\milli\liter$ } \\
\hline
\sainz & & MOI: 0.01 $\ffu$/cell & \multirow{4}{\txcolwi}{\raggedright Extracellular HCV kinetics during multiple-cycle infection. Used to fix $\bar{R} = \unit{120}{\frac{HCV}{cell}}$ and \useabm.}\\
Fig.\ 3B (open diamond) & Fig.\ \ref{freprod}B (red stars) & Virus: JFH1  & \multirow{4}{\txcolwi}{} \\
Fig.\ 4A (bars) & Fig.\ \ref{freprod}B (red circles) & Cell line: Huh7  & \multirow{4}{\txcolwi}{} \\
 & & & \\
\hline
\multicolumn{4}{l}{\multirow{1}{\textwidth}{\raggedright Number of cells infected at \unit{72}{\hpi} in the presence of nAbs, expressed relative to control}} \\
\hline
\timpe & & MOI: 0.01$^b$ & \multirow{4}{\txcolwi}{\raggedright Number of cells infected in presence of nAbs is (43$\pm$16)\% that for control. Used to \useabm.}\\
Fig.\ 2D (grey bar) & Fig.\ \ref{freprod}C (red curve) & Virus: JFH1  & \multirow{5}{\txcolwi}{} \\
 &  & Cell line: Huh7.5  & \multirow{5}{\txcolwi}{} \\
 & & & \\
\hline
\hline
\multicolumn{4}{l}{\raggedright Foci size distribution data} \\
\hline
\timpe & & MOI: 0.01$^b$ & \multirow{4}{\txcolwi}{\raggedright Provides estimate for the foci size distribution at $\unit{72}{\hpi}$. Used to validate ABM results post PPLD estimation.} \\
Fig.\ 1E,F (black bars) & Fig.\ \ref{ftimpeclouds} & Virus: JFH1  & \multirow{4}{\txcolwi}{} \\
 &  & Cell line: Huh7.5  & \multirow{4}{\txcolwi}{} \\
 & & & \\
\hline
\hline
\end{tabular}
$^a$ JFH-m4 is HCV JFH1-derived, but harbours mutations in E2 and p7 proteins (see Discussion for details). \\
$^b$ Units of the MOI were not specified by the authors.
\end{table}

\begin{figure}
\begin{center}
\resizebox{0.4\textwidth}{!}{\includegraphics{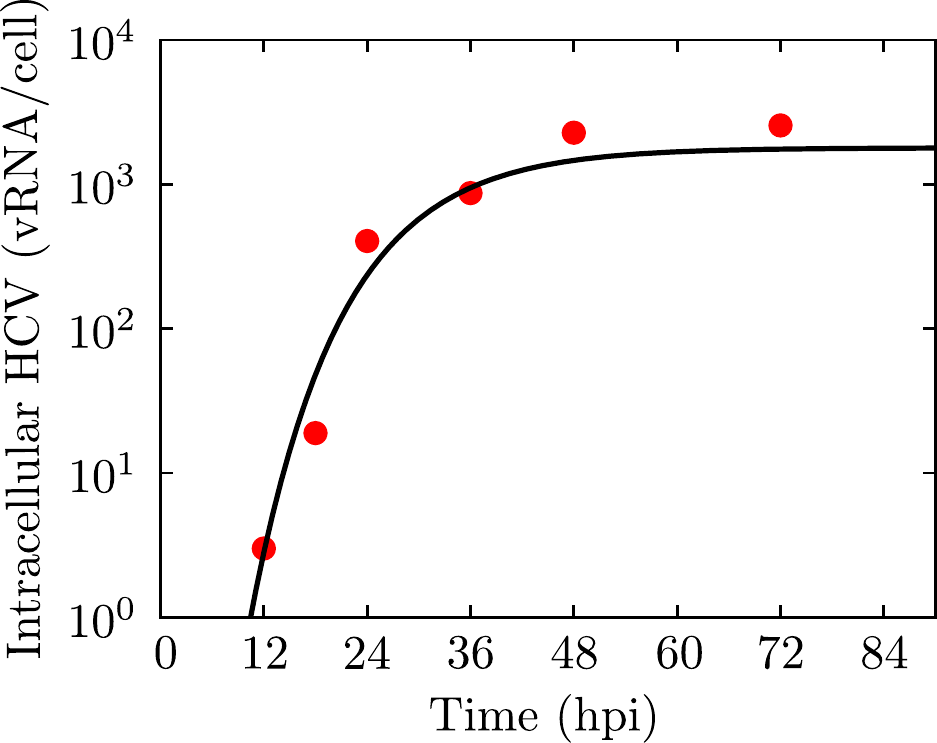}}
\caption{\textbf{Best-fit to estimate the intracellular HCV RNA growth rate.} %
The data set by \keum (Fig 1A therein) consists in the mean over three independent experiments of the concentration of positive-strand intracellular HCV (vRNA/cell) over the course of an infection with HCV JFH-m4 in Huh7.5.1 cells in vitro at a MOI of \unit{6}{FFU/cell}. Fitting Eqn.\ \eqref{intraRNA} (line) to the data (circles) enabled us to estimate the growth rate of intracellular HCV RNA to be $\alpha = \unit{0.097}{\perh}$ (see Methods for details).
}
\label{fkeum12}
\end{center}
\end{figure}

The data from \keum were used to parameterize the single-cell, intracellular HCV replication kinetics described by Eqn.\ \eqref{intraRNA}. The data consist of the mean intracellular, positive-strand HCV RNA per cell, measured over the course of three independent in vitro experiments in which Huh7.5.1 cells were infected at a high MOI ($\unit{6}{FFU/cell}$). The high MOI serves to roughly synchronize the time of infection of all hepatocytes in the assay such that the time course of average intracellular HCV within the culture as a whole represents that within a single HCV-infected cell. We fit the Gompertz curve, Eqn.\ \eqref{intraRNA}, representing the intracellular HCV RNA replication, to the \keum data, as shown in Figure \ref{fkeum12}, with details provided in Methods. From this, we obtained an estimate for the intracellular HCV growth rate of $\alpha = \unit{0.097}{\perh}$.

\begin{figure} 
\resizebox{\textwidth}{!}{\includegraphics{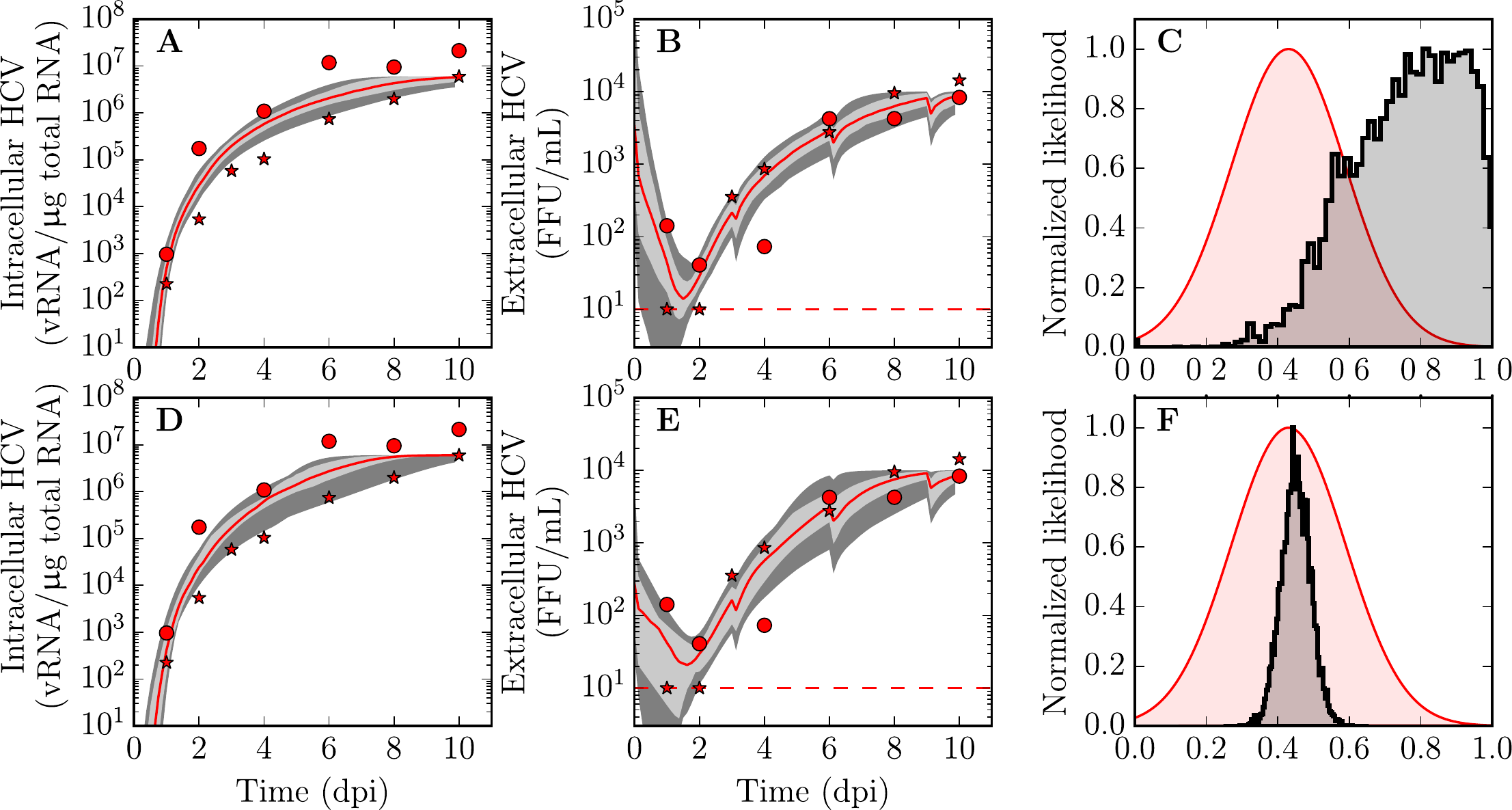}}
\caption{%
\textbf{Comparison of experimental vs ABM-generated course of HCV infection in vitro.} %
The time course of \textbf{(A)} intracellular HCV RNA and \textbf{(B)} extracellular infectious HCV, measured in duplicate (red stars and circles) from \sainz are shown against the median of ABM-simulated time courses in the absence of nAbs (red line) and its associated 68\% and 95\% percentiles of the $y$-value at each $x$-value (time point) for 500 ABM simulations performed (grey bands). Data points from \sainz on the red dash line in panel (B) are at the limit of detection (see Methods for discussion). \textbf{(C)} The relative percentage of HCV-infected hepatocytes in the presence vs absence of anti-HCV nAbs at \unit{72}{\hpi} of ($43\pm16$)\% reported by \timpe (red distribution) against the PPLD for that measure predicted by the ABM (black histogram). (\textbf{D,E,F}) Same as (A,B,C), but results from a different MCMC PPLD estimation wherein the weight (importance) of the \timpe measurement (C,F) was increased by decreasing $\sigma_f$ (in Eqn.\ \eqref{elikeli}) from 0.16 in (A,B,C) to 0.04 in (D,E,F). The drop in ABM-simulated extracellular HCV titer recurring every 3 days in panel (B,E) simulates the replacement of the extracellular medium which took place over the course of the experimental infection. See Methods for details on how the ABM-predictions were performed.}
\label{freprod}
\end{figure}

The data from \sainz were used to parametrize the kinetics of HCV dissemination through the cell culture. To arrest proliferation of hepatocytes during infection, \sainz treated Huh7 cells with dimethyl sulphoxide (DMSO) for 20 days prior to HCV infection. The DMSO-treated Huh7 cells were infected with an inoculum of HCV JFH1 at a low MOI (\unit{0.01}{\ffu/cell}), resulting in several cycles of infection which enabled us to establish the kinetics of infection spread, unobserved in \keum. The data consist in frequent measurements of both the concentration of intracellular HCV RNA quantified via RT-qPCR (\Ru) shown in Figure \ref{freprod}A, and the concentration of extracellular infectious HCV in the supernatant quantified via titration on Huh7 cells immunostained with anti-HCV E2 Abs reported as focus forming units (\Vu) shown in Figure \ref{freprod}B. While this data set provides information about the kinetic of HCV dissemination over multiple cycles of infection, it does not provide spatial data nor information about the mode of dissemination.

In order to specifically address the relative contribution of the free virus versus cell-to-cell dissemination of HCV in vitro, additional data from \timpe were considered in parametrizing the ABM. \timpe infected Huh7.5 cells with JFH1 at a MOI of 0.01 (presumably \unit{0.01}{\ffu/cell}, though this is not specified in \cite{timpe08}) for one hour. The infections were performed in the presence of either a control (anti-dengue nAbs) or anti-HCV nAbs which were shown therein to neutralize $>95\%$ of cell-free, extracellular HCV. The nAbs were added to the extracellular medium at \unit{8}{\hour} post-infection (\hpi), and replenished every \unit{24}{\hour} thereafter to maintain the neutralizing activity. At $\unit{72}{\hpi}$, the number of HCV-infected hepatocytes resulting from infection in the presence of anti-HCV nAbs and the control infection were counted. Unfortunately, \timpe do not report any kinetic information (measurements of virus or cell infection over time) from the time course of these infections. Instead, they provide the number of infected hepatocytes in the presence of anti-HCV nAbs and in the control infection, at \unit{72}{\hpi}, in the form of a bar graph (Fig.\ 2D in \cite{timpe08}), with error bars representing the standard error of the mean (SEM) based on 4 independent experiments. From this bar graph, we extract (see Methods) a single quantity: at \unit{72}{\hpi}, the number of HCV-infected hepatocytes in the presence of anti-HCV nAbs is ($43\pm16$)\% of that in the control experiment. This single quantity is represented in Figure \ref{freprod}C as a Gaussian probability density function with a mean of $43\%$ and a standard deviation of $16\%$.

From the infection kinetics in the \sainz data (Figure \ref{freprod}A,B), some of the ABM parameters can be determined. For example, the extracellular infectious HCV concentration reaches a steady state of $\bar{V}=\unit{10^4}{\Vu}$, and intracellular HCV RNA reaches a steady state of \unit{6\times10^6}{\hcvr} copies per $\micro\gram$ of total cellular RNA. Assuming that all hepatocytes are infected at steady state, and one hepatocyte contains \unit{20}{\pico\gram} (or \unit{20\times10^{-6}}{\micro\gram}) of total cellular RNA \cite{wilkening03, detomassi02}, then each HCV infected Huh7 cell in the \sainz experiment contained on average approximately $\bar{R} = \unit{120}{HCV}$ RNA copies at steady state. Having fixed $\alpha$, $\bar{V}$ and $\bar{R}$, only 4 biological parameters (\trnaoffset, $c$, $\beta_V$, $\beta_C$) remain to be determined. In addition, two parameters are required to describe the experimental conditions: the initial virus inoculum $V(t=0)$, and the efficacy with which this inoculum is rinsed at the end of the incubation period, $0 \le \frinse \le 1$.

For each set of ABM parameters considered, (\trnaoffset, $c$, $\beta_V$, $\beta_C$, $V(0)$, \frinse), the course of HCV infections is simulated twice: once in the presence of anti-HCV nAbs and again in their absence. To simulate the infection in the presence of nAbs, the extracellular infectious HCV concentration, $V(t)$, and rate of HCV release, $p_V(t)$, are set to 0 at $t\ge\unit{8}{\hpi}$, such that $V(t\ge\unit{8}{\hpi})=0$. The time course for intracellular and extracellular HCV in the absence of nAbs ($R(t),V(t)$, respectively) is compared against the kinetic \sainz data sets, and the number of infected cells in the presence divided by that in the absence of nAbs at \unit{72}{\hpi} in the ABM-simulated infections, \fI (a single number), is compared against the expected value of $43\%$ from \timpe. Since these three sets of measurements ($R(t),V(t),\fI$) have different number of points, units, and standard deviations, the sum-of-squared residuals (SSR) for each measurement set were normalized by the variance of their respective residuals, before being summed together, as described in Methods, Eqn.\ \eqref{elikeli}. The time course for $R$ and $V$ in the presence of nAbs, which was simulated by our ABM in order to determine \fI, is not used because these data were unfortunately not measured experimentally.

Because the ABM-simulated infections are stochastic, one set of parameters, (\trnaoffset, $c$, $\beta_V$, $\beta_C$, $V(0)$, \frinse), does not map to a unique SSR value. To address this issue, we introduce one additional nuisance parameter, the random number seed (RNS), used to seed the pseudo random number generator at the start of each ABM run. This renders the now 7-dimensional parameter space defined by (\trnaoffset, $c$, $\beta_V$, $\beta_C$, $V(0)$, \frinse, RNS) deterministic, and yields a single goodness-of-fit measure, $\ssr(\vec{p})$, for any given parameter set $\vec{p}$, as specified in Methods, Eqn.\ \eqref{elikeli}. To estimate $\vec{p}$, we used a Markov chain Monte Carlo (MCMC) approach, as described previously \cite{paradis15,simon16,beauchemin17}. Briefly, each candidate parameter set, $\vec{p}$, is accepted with a probability based on its Bayesian likelihood, defined in Eqn.\ \eqref{elikeli} in Methods. The collection of all accepted parameter sets forms the posterior parameter likelihood distributions (PPLDs) whose modes and 95\% credible intervals (CI) are reported in Table \ref{param-table}.

\begin{table}
\begin{center}
\caption{MCMC-estimated PPLDs for the ABM.}
\label{param-table}
\begin{tabular}{ll}
 & Mode [95\%CI]$^a$ \\ [0.2ex]
\hline
\multicolumn{2}{c}{--- Fixed or computed directly from data (see footnotes) ---} \\
Number of hepatocytes, $N_\text{cell}$ (cell) &
	10,000 $^b$ \\
Intracellular HCV RNA growth rate, $\alpha$ (\perh) &
	$0.097$ $^c$\\
Equilibrium extracellular infectious HCV concentration, $\bar{V}$ (\Vu) &
	$10^4$ $^d$\\
Equilibrium intracellular HCV RNA concentration, $\bar{R}$ (HCV RNA/cell) &
	$120$ $^d$\\
\hline
\multicolumn{2}{c}{--- PPLDs estimated via MCMC ---} \\
Intracellular HCV RNA replication delay, \trnaoffset (\hour) &
	$19$ $[9.2,26]$ \\
Rate of loss of HCV infectivity, $c$ (\perh) &
	$0.27$ $[0.049,0.30]$ \\
Rate of infection via cell-to-cell, $\beta_C$ (\perh) &
	$85$ $[3.6,100]$ \\
Initial virus concentration, \Vo (\Vu) &
	$10^{3.7\ [2.0,5.5]}$ \\
Multiplicity of infection, MOI (\ffu/cell) &
	$10^{-2.9\ [-3.8,-2.0]}$ \\
Effectiveness of the rinse, \frinse &
	$0.47$ $[0.036,1.0]$ \\
\hline
\multicolumn{2}{c}{--- Computed from MCMC PPLDs (see Methods) ---} \\
Rate of infection via free virus, $\beta_V$ (\perh) &
	$10^{-6.5\ [-8.2,-4.9]}$ \\
Percent infected cells w/wo nAbs at \unit{72}{\hpi}, $f_I(\unit{72}{\hour})$ (\%) &
	$91$ $[49,100]$ \\
Percent of infections via cell-to-cell at \unit{10}{dpi} (\%) &
	$99$ $[84,100]$ \\
Percent of infections via free virus at \unit{10}{dpi} (\%) &
	$1.2$ $[0.038,16]$ \\
\hline
\end{tabular}
\end{center}

\vspace{-0.5em}
$^a$ Mode and 95\% CI determined from the PPLDs.\\
$^b$ The spatial extent of the hexagonal ABM is (100$\times$100) hepatocytes or $\sim$\unit{(2\times2)}{\milli\metre} assuming one hepatocyte is $\sim$\unit{20}{\micro\metre} in diameter \cite{lohmann99}. \\
$^c$ Estimated from the \keum data (see Methods and Figure \ref{fkeum12}).\\
$^d$ Steady states of extracellular infectious HCV and intracellular HCV RNA in experimental data from \sainz.
\end{table}

Figure \ref{freprod}(A,B) presents the intracellular HCV RNA and extracellular infectious HCV kinetics over the course of the parametrized ABM-simulated HCV infections in the absence of nAbs against the \sainz experimental data. The MCMC-parametrized ABM reproduces this HCV infection kinetics well. In Figure \ref{freprod}C, however, the PPLD for the number of HCV-infected hepatocytes in the presence of nAbs relative to that in the absence of nAbs at \unit{72}{\hpi} in our parametrized ABM-simulated infections visually does not appear to provide a good agreement with that reported in \timpe. Quantitatively, this disagreement is not statistically significant, yielding a fold-change of 1.7 (0.73--5.2, 95\% CI), i.e.\ 1 is included within the 95\% CI of the fold-change. The disagreement is more apparent in Figure \ref{freprod}C because it shows a single measure or point, but worse agreements can be found at other single time point measurements in Figure \ref{freprod}(A,B), e.g.\ at \unit{4}{dpi} in Figure \ref{freprod}A. In establishing the parameter likelihood for the MCMC method, namely Eqn.\ \eqref{elikeli} in Methods, each measure is weighted by its standard deviation, i.e.\ the degree of uncertainty in that measure, such that less emphasis is placed on measures with larger uncertainty, as it should be. This should not be interpreted to mean that the ABM could not simultaneously fit the kinetic data a little less well so as to provide a better agreement with the measure shown in Figure \ref{freprod}C if this weighting was adjusted.

To demonstrate this, we placed $16\times$ more weight ($\sigma_f^2$ in Eqn.\ \eqref{elikeli} decreased from $0.16^2$ to $0.04^2$) on the measure from \timpe, and repeated the MCMC PPLD estimation. The results are shown in Figure \ref{freprod}(D,E,F), and the modes and 95\% CI of the corresponding PPLDs are in Table \ref{tmoretimpe}. As expected, the variability in the infection kinetics curves (Figure \ref{freprod}A,B vs D,E) increases slightly while the relative percentage of HCV-infected cells in the presence vs absence of nAbs (Figure \ref{freprod}C vs F) now provides perfect agreement with the measure reported by \timpe, albeit with an unphysically narrow PPLD standard deviation. Importantly, imposing this unrealistically strict constraint for agreement with the \timpe measure causes a statistically significant change in said measure only ($1 \not\in [0.40,0.98]$, 95\% CI of fold-change), and not in any other parameters or computed quantities (see Methods).

\subsection{Determining the modes of HCV dissemination}

\begin{figure}
\resizebox{\textwidth}{!}{\includegraphics{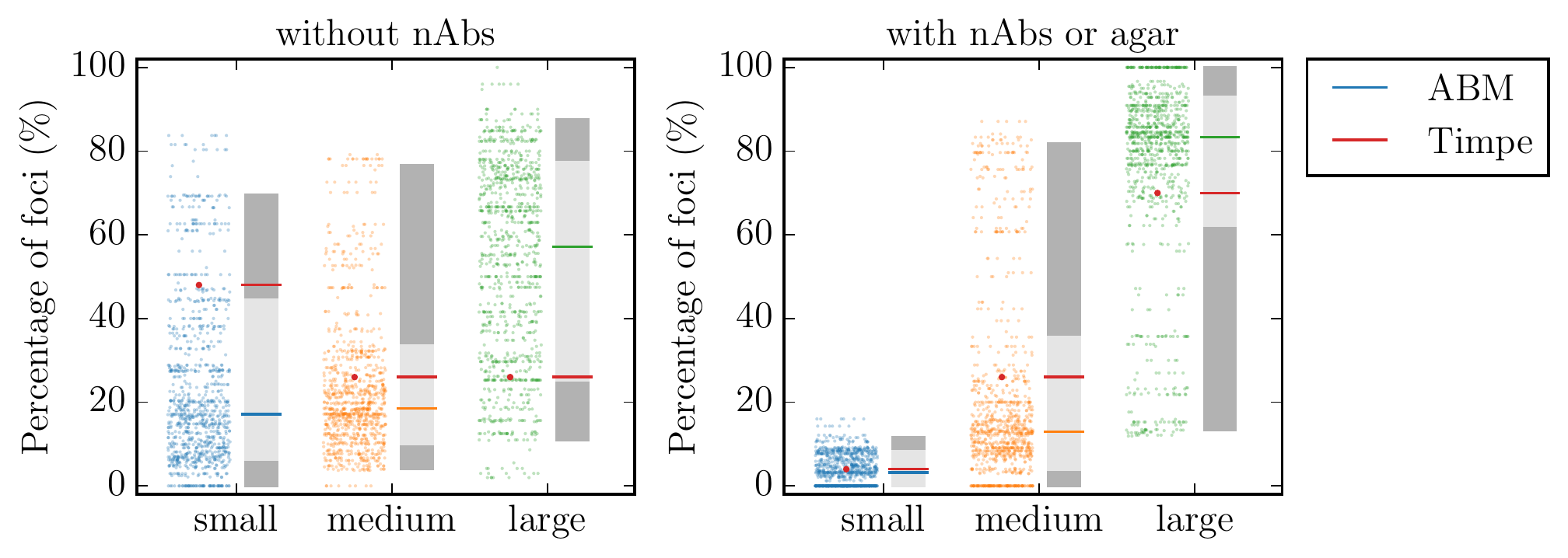}}
\caption{%
\textbf{Quantitative comparison of ABM-simulated and experimental foci size distributions.} The ABM-simulated percentage of small ($<10$ cells), medium (10--50 cells), and large ($>50$ cells) foci at \unit{72}{hpi} for 1000 simulations, each simulation containing between 1 and 455 foci, are shown both as scatter dots and percentile bars (median, 68\% CI, 95\% CI = bar, light grey, dark grey). These are compared against that observed by \timpe (Fig.\ 1E,F therein) who counted and categorized 50 foci (single red scatter dot and red bar). Note that \timpe do not specify if their 50 foci were counted within the same well or experiment, or summed over several independent experiments, nor provide any error bars. Results are shown for HCV infections in the absence (left) and presence (right) of free virus dissemination, which was blocked in our ABM by adding nAbs, whereas \timpe used agar. The ABM-simulated foci size distributions are in good agreement with those observed by \timpe (agrees within the 68\% CI in all but one that agrees within the 95\% CI). See Methods for details on how the ABM-predictions were performed. 
}\label{ftimpeclouds}
\end{figure}

Figure \ref{ftimpeclouds} shows that the parametrized ABM's simulated foci size distributions agree well with those reported in \timpe in the presence and absence of nAbs (or agar). This provides some degree of validation of the spatial predictions of the parametrized ABM.

Next, the parametrized ABM was used to run in silico simulations to robustly quantify and establish the percentage of hepatocytes infected via each infection route: cell-to-cell and free virus. To do so, we allowed each simulated infection parametrized from our PPLD to run their full course (up to 10 days post-infection) such that all or almost all hepatocytes have become infected. At the end of each of these simulated infections performed in the absence of nAbs, we recorded the percentage of all infected hepatocytes infected via the cell-to-cell and free virus route, reported in Table \ref{param-table}. The parametrized ABM estimates that only about 1.2\% (0.038\%--16\%, 95\% CI) of all infection events over the full course of the infection (10 days) in the absence of nAbs occur via the free virus route, with the remainder ($\sim$99\%) from direct cell-to-cell infection. This estimate is not significantly affected by placing more weight on the measure from \timpe, which yields 2.1\% (0.25\%--30\%, 95\% CI). It might seem counter-intuitive that blocking 1\% of infections occurring via the free virus route using nAbs would result in the staggering 57\% reduction in the number of cells infected reported by \timpe, but there are two main reasons for this seeming discrepancy. 

\begin{figure}
\resizebox{\textwidth}{!}{\includegraphics{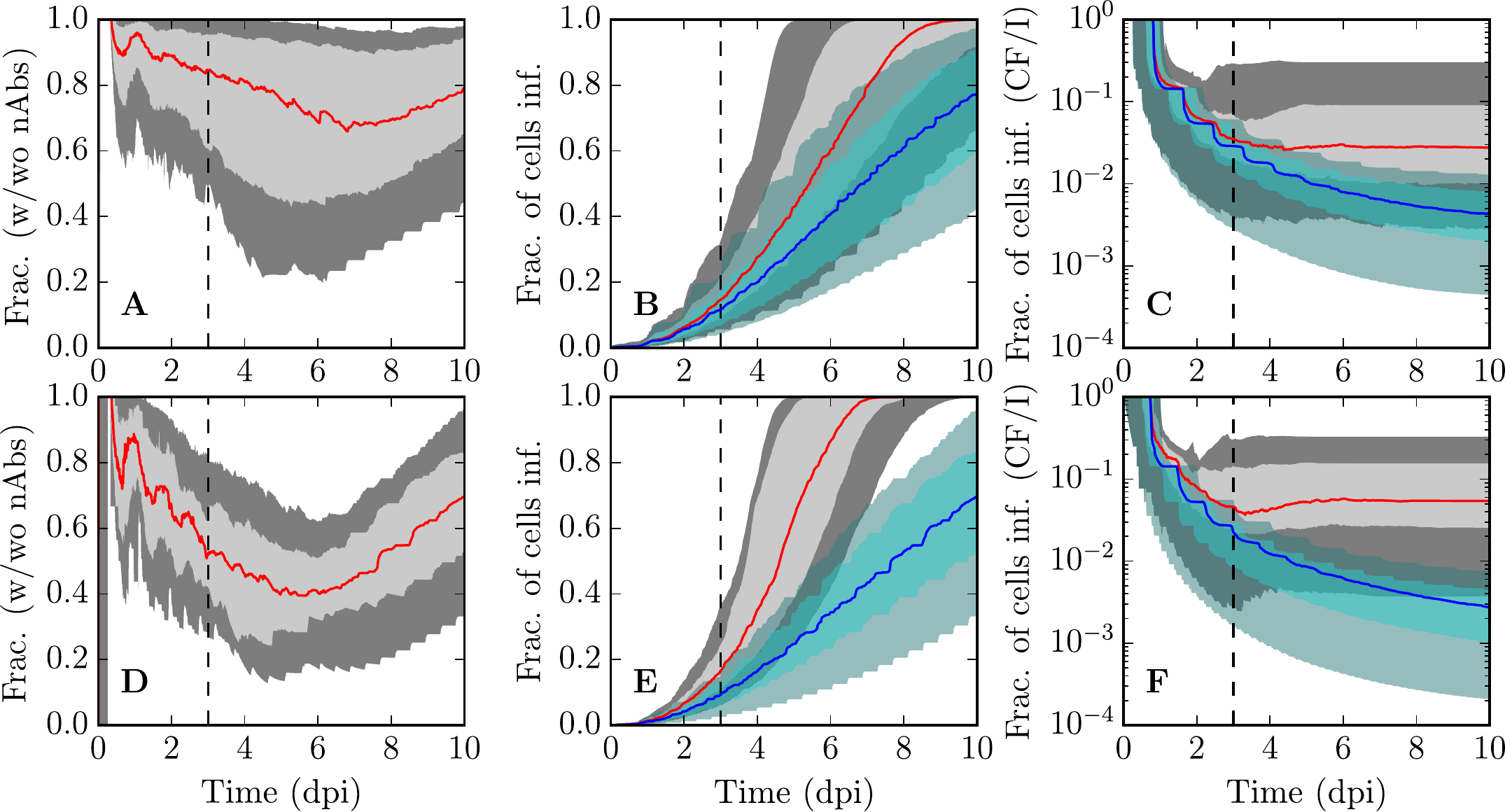}}
\caption{\textbf{Fraction of infection by each mode, and the relative impact of nAbs.} ABM-simulated time courses for the \textbf{(A)} fraction of cells that are HCV-infected in the presence of nAbs over that in their absence; \textbf{(B)} fraction of cells that are HCV-infected in the absence (grey) and presence (blue) of nAbs; and \textbf{(C)} fraction of HCV-infected cells that were infected via free virus (CF) over the fraction of all cells that are HCV-infected (I). The solid line (red or blue) indicate the median, and the pale and darker bands (grey or cyan) indicate the 68\% and 95\% percentiles of the $y$-value at each $x$-value (time point) for 500 ABM simulations performed (see Methods). \textbf{(D,E,F)} same as (A,B,C), but results from the different MCMC PPLD estimation wherein the weight of the \timpe data was increased ($\sigma_f:0.16 \rightarrow 0.04$ in Eqn.\ \eqref{elikeli}). The vertical dashed line at \unit{3}{dpi} illustrates the time of the fraction of cells recorded by \timpe.
}
\label{ffractiming}
\end{figure}

One reason is that \unit{72}{hpi} is still relatively early in the infection process. This is illustrated in Figure \ref{ffractiming} which shows the ABM-simulated time course for various measures of the fraction of infected cells for corresponding intra- and extra-cellular HCV concentrations reported in Figure \ref{freprod}. For example, Figure \ref{ffractiming}B shows that by \unit{72}{hpi}, only $\sim$13\% of hepatocytes in our ABM-simulated infections have become infected. Since experimental infections are initiated by a free virus inoculum, the fraction of infections due to free virus is initially 100\% of infection events. But as time progresses, and several rounds of the more rapid cell-to-cell infections take place, the proportion of infections in the presence vs absence of nAbs will progressively decrease. In other words, the 43\% reported by \timpe should not be considered a definitive quantification of the contribution of each mode, but rather a much less informative ``snap-shot'' of that quantity at a somewhat arbitrary time post-inoculation.  

\begin{figure}
\resizebox{0.32\textwidth}{!}{\includegraphics{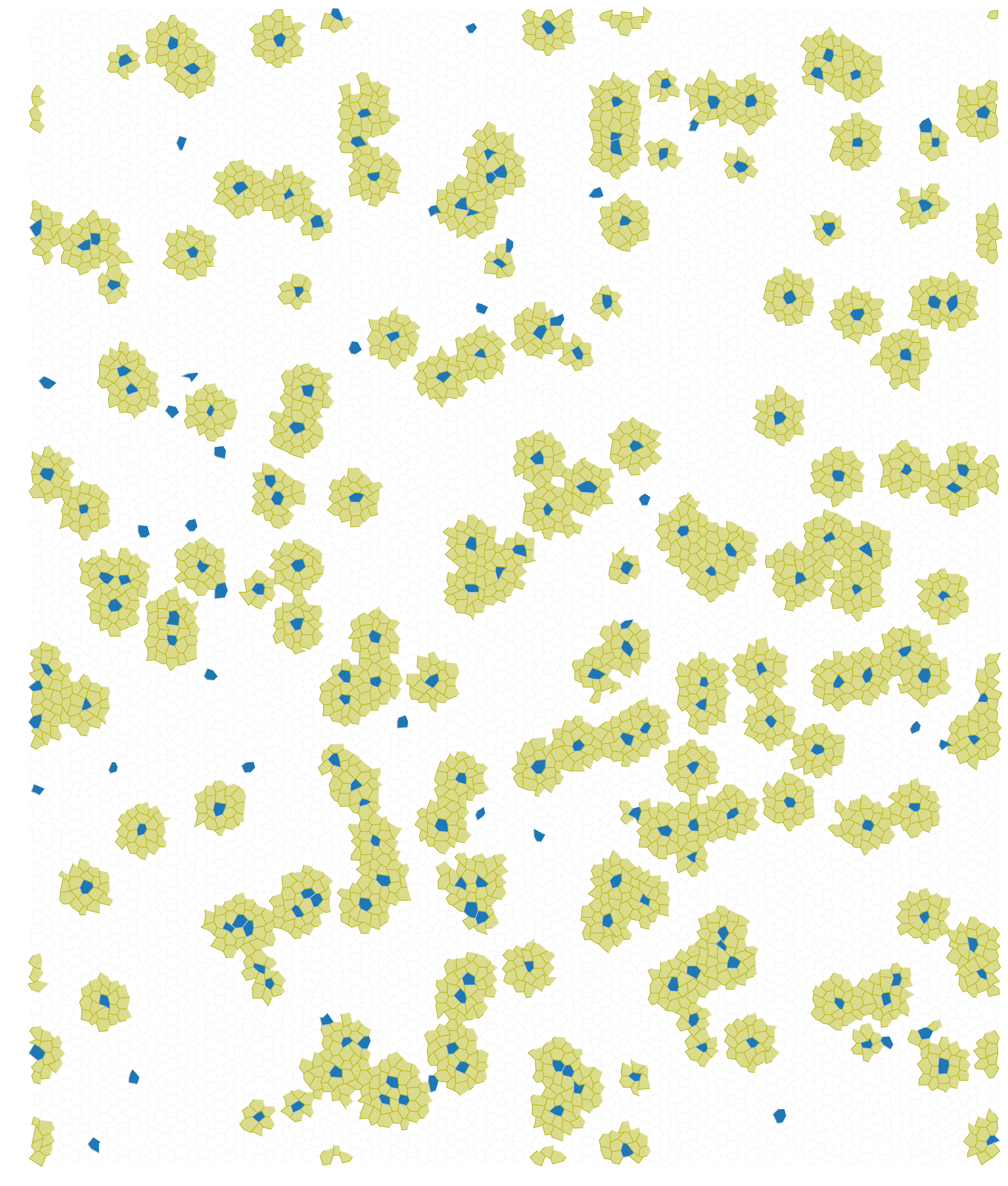}}
\resizebox{0.32\textwidth}{!}{\includegraphics{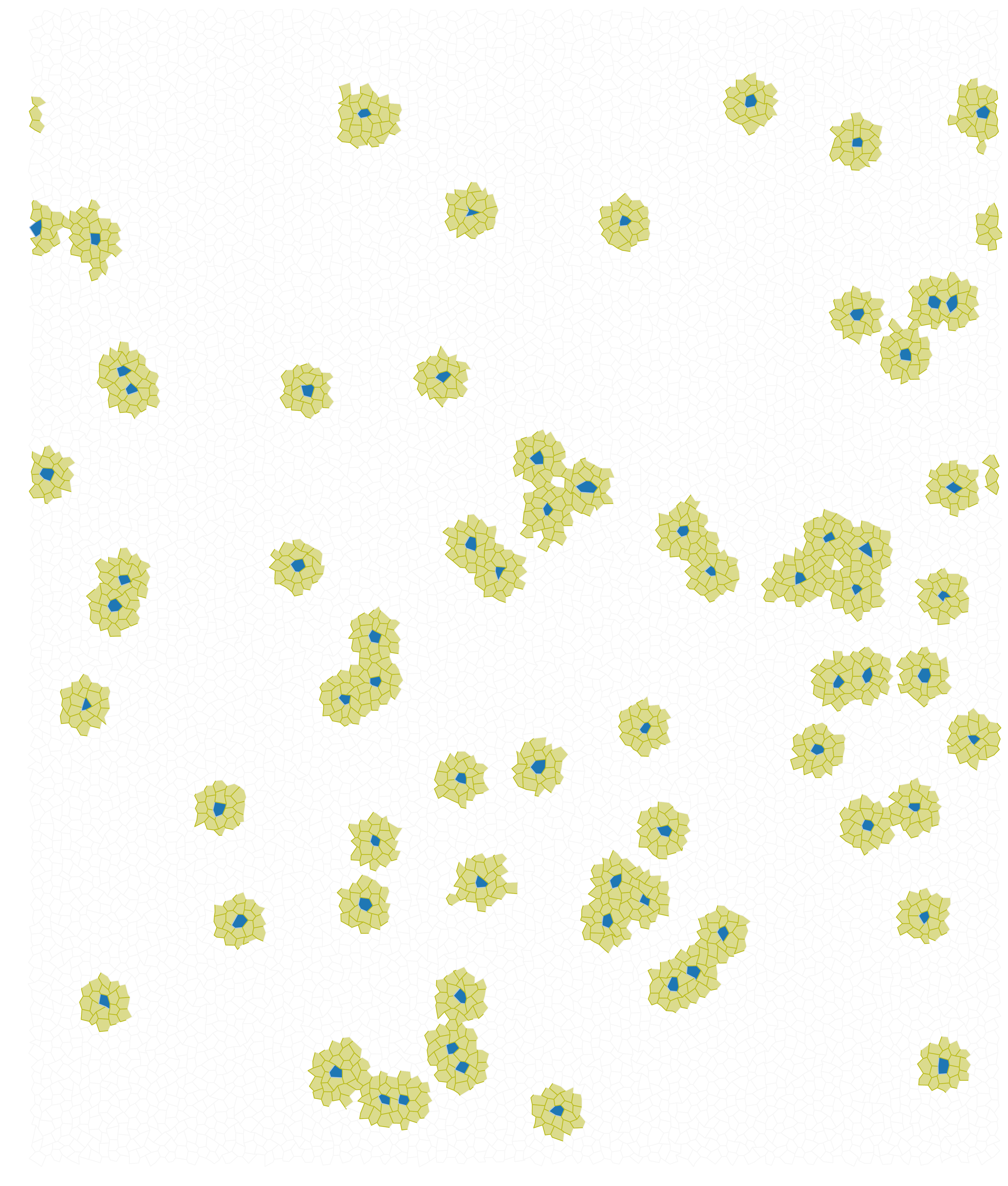}}
\resizebox{0.32\textwidth}{!}{\includegraphics{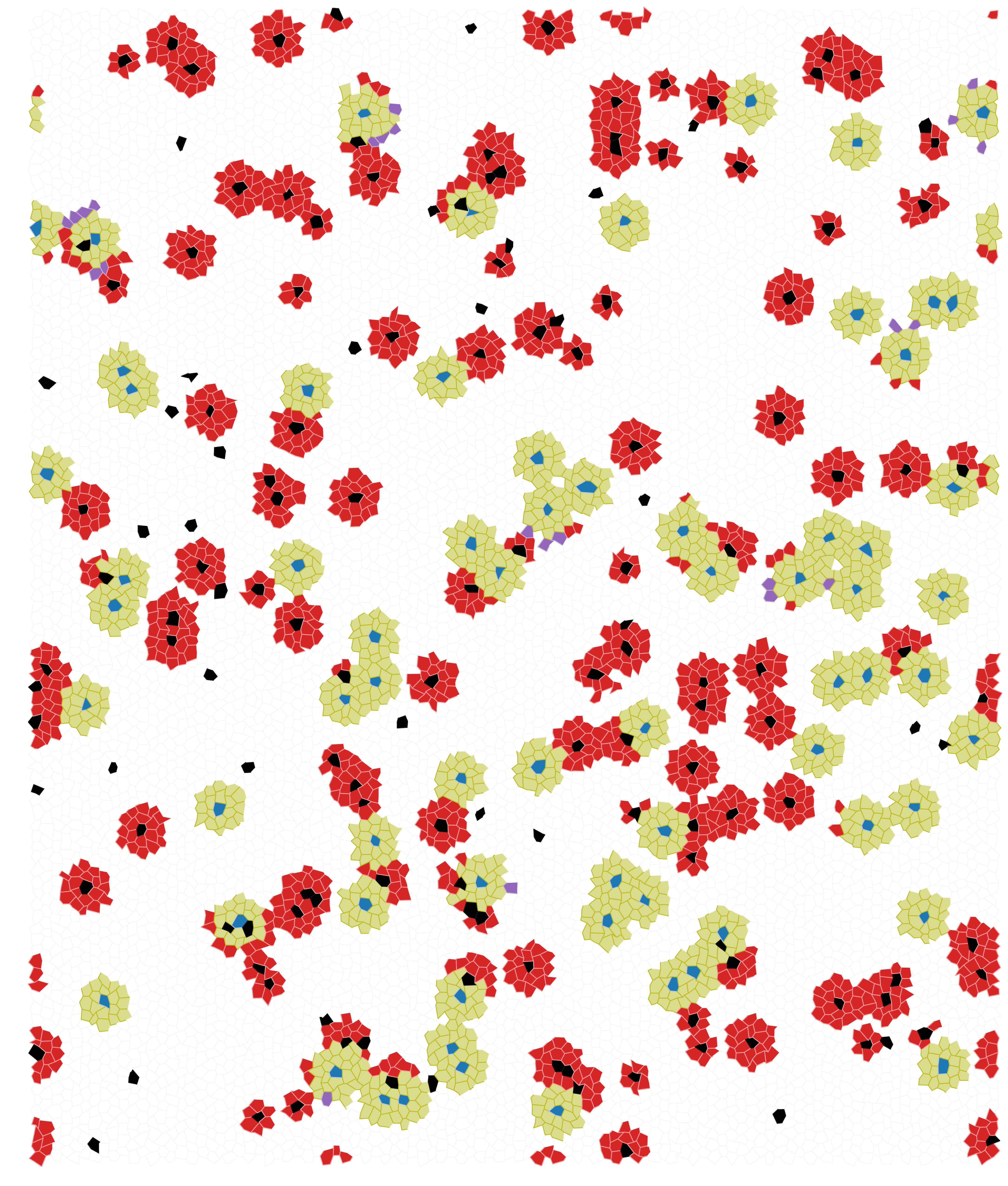}}
\caption{\textbf{Visualizing the spatial spread of an ABM-simulated in silico HCV infection.} Spatial representation of an ABM simulated in silico HCV infection at \unit{72}{hpi}, using a parameter set pulled at random from the PPLD, (left) without nAbs, (centre) with nAbs and (right) difference between the two. Uninfected hepatocytes (white) become infected via free virus (blue) or by spread of infection via cell-to-cell (yellow). In the right panel, cells which were infected in the simulation without nAbs, but are no longer infected when nAbs were added appear in black for free virus infected cells, and red from cell-to-cell infected cells. In rare cases, the simulation with nAbs has some cells infected which were not infected without nAbs (purple) because, despite the simulation with and without nAbs having been run with the same random number seed, different lists of random numbers were generated in each run. This visual comparison shows that the removal of free virus by nAbs not only affects the number of hepatocytes infected by free virus but also affects the number infected via cell-to-cell transmission as each new focus begins with a hepatocyte infected by free virus and then grows almost exclusively via direct, local, cell-to-cell infection.}
\label{fsshot}
\end{figure}

The other reason is that there is a synergy between infections via the free virus and cell-to-cell routes, such that a small reduction in the number of free virus infection greatly reduces the number of cell-to-cell infections. This can be seen in Figure \ref{fsshot} which presents a visualization of the spatial spread of an ABM-simulated in silico HCV infection at \unit{72}{hpi}. Each new focus is initiated by a seed hepatocyte infected distally via free virus (blue), claiming a new, fresh, uninfected area. Direct cell-to-cell infection (yellow) then takes over to efficiently disseminate infection from the initial seed to its neighbouring hepatocytes, which go on to directly infect their neighbours, and so on. The in silico labelling of the mode by which each hepatocyte was infected highlights that indeed most hepatocytes are infected via direct cell-to-cell contact (yellow) rather than via free virus (blue). In particular, it shows that at \unit{72}{hpi}, the largest and thus oldest foci have 2--3 ``concentric rings'' of infected hepatocytes, which add up to 20--40 hepatocytes, of which only one was infected via free virus, and the remainder via cell-to-cell, which corresponds to $\sim$3\% (1/30) of infections proceeding via the free virus route.

\begin{figure}
\resizebox{\textwidth}{!}{\includegraphics{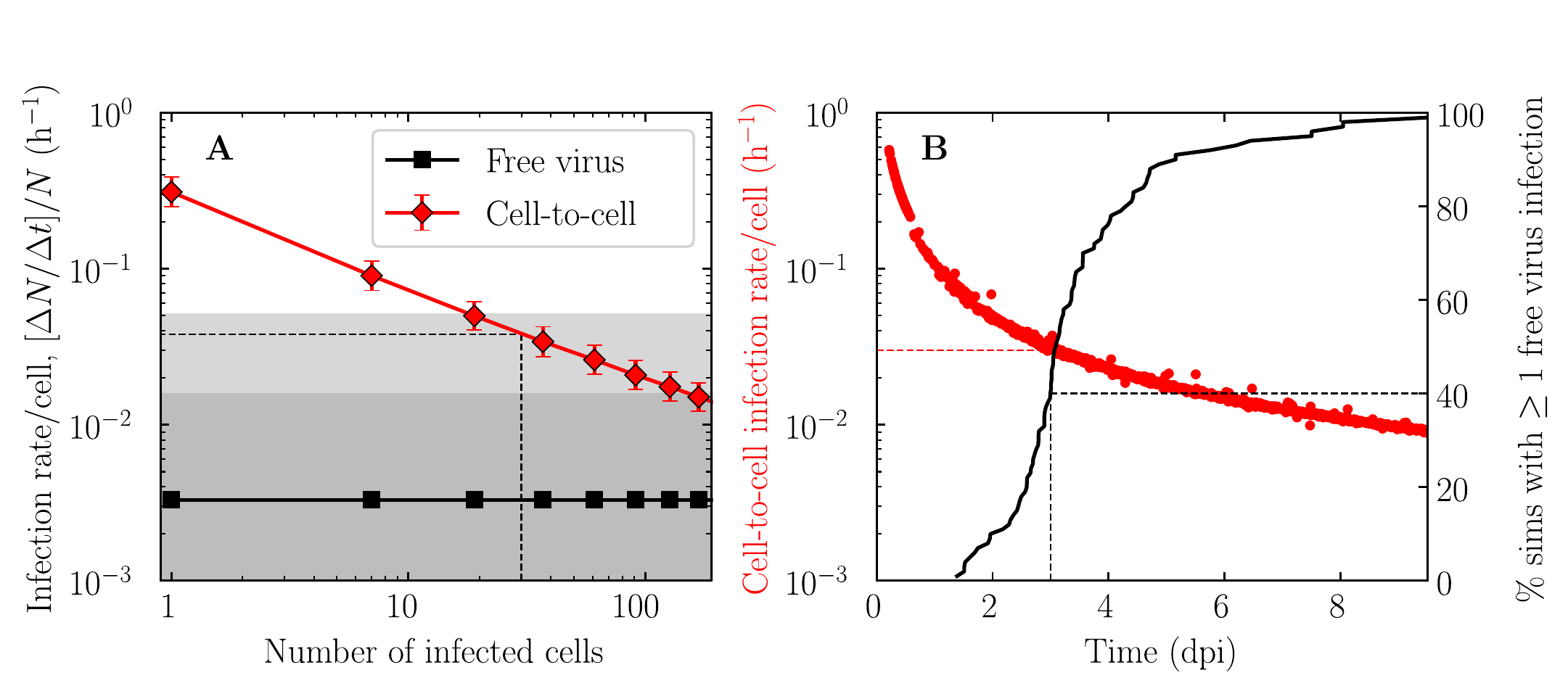}}
\caption{%
\textbf{HCV kinetics for dissemination via cell-to-cell alone or free-virus alone.} The ABM was used to simulate the course of 100 infections, using parameter sets pulled at random with replacement from the PPLD, initiated with a single infected hepatocyte wherein infection disseminates via either only cell-to-cell (red, $p_V(t)=0$) or only free virus (black, $\beta_C=0$). \textbf{(A)} The infection rate per cell as a function of the number of infected cells remains constant for dissemination via free virus (black), but decreases for cell-to-cell dissemination (red). The solid lines are the geometric means, the error bars for cell-to-cell infections are the geometric standard deviations, and the pale and darker grey bands for free virus infections are the 68\% and 95\% CI over the simulations performed. Dash lines show that when a single focus has grown to $\sim$2--3 ``concentric rings'' of infected cells (20--40 infected cells), the cell-to-cell infection rate agrees within the 95\% CI of the free virus infection rate. \textbf{(B)} The decrease in the cell-to-cell infection rate per cell as a function of time for cell-to-cell dissemination alone (red, left $y$-axis) is shown against the \% of ABM simulations in which a single growing focus disseminating through cell-to-cell infection alone would have caused at least one infection via the free virus route (black, right $y$-axis). Dash lines show that at \unit{3}{dpi}, when the largest foci have a 2--3 ``concentric rings'' of infected cells, there is a $\sim$40\% probability that a new focus will have been initiated via free virus and the cell-to-cell infection rate per cell has decreased to $\sim$5\% of its value from the start of the infection.}\label{fc2cvsfv}
\end{figure}

A more quantitative analysis of this principle is presented in Figure \ref{fc2cvsfv} where the ABM was used to simulate infections starting from a single infected seed hepatocyte, progressing exclusively via either free virus or cell-to-cell dissemination. As the single focus grows from its initial seed via cell-to-cell infection, the decreasing perimeter-to-area ratio of the focus leads to decreasing number of infections caused by an increasing number of infected hepatocytes. In contrast, when infection proceeds via free virus alone, the rate of infection per infected hepatocyte remains constant. Figure \ref{fc2cvsfv}A shows that once a single focus has grown to $\sim$2--3 ``concentric rings'' of infected hepatocytes (20--40 hepatocytes), the cell-to-cell infection rate per infected hepatocyte approaches that for free virus infections (agrees within its 95\% CI). If, at that point, a new focus is initiated via free virus, the effective cell-to-cell infection rate per infected hepatocyte in that new focus is ``reset'' to its maximum value, speeding up the overall rate at which new hepatocytes become infected. Given that Figure \ref{fsshot} shows that the largest foci have a 2--3 ``concentric rings'' of infected hepatocytes at \unit{72}{hpi}, Figure \ref{fc2cvsfv}B shows that by that time, there is a $\sim$40\% probability that a new focus will have been initiated via free virus and the cell-to-cell infection rate per cell has decreased to $\sim$5\% of its value from the start of the infection.

\cleardoublepage
\section{Discussion}

The hepatitis C virus (HCV) is one of many virus which take advantage of both the cell-free and cell-to-cell mechanisms of transmission to spread effectively \cite{mothes10}. Herein, we introduce a spatial ABM simulator for HCV infection spread in vitro. In the ABM, the in vitro infection system is represented as a two-dimensional hexagonal grid where each grid site corresponds to one hepatocyte. The ABM tracks both the overall extracellular infectious HCV concentration in the supernatant, and the amount of intracellular HCV RNA within each hepatocyte. It captures both transmission via free virus in the supernatant, and via direct cell-to-cell transmission to a hepatocyte's six immediate neighbours. Published experimental data from HCV infections performed in vitro in the presence and absence of nAbs were used to determine the unknown kinetic parameters of the ABM. In particular, data from in vitro infections wherein HCV dissemination via free virus is suppressed by nAbs were considered. Once calibrated, the ABM was used to simulate the full course of an in vitro HCV infection, and to identify and tag the mode of HCV transmission responsible for the infection of each hepatocyte. Through this process, it was determined that 99\% (84\%--100\%, 95\% CI) of all hepatocytes are infected via the cell-to-cell route, with distal, free virus infection accounting for $\sim$1\% of hepatocyte infections. This is the first time this quantity has ever been determined for HCV infections in vitro.

Some of the experimental data used herein to calibrate the ABM were that of \timpe which performed in vitro HCV infections in the presence and absence of anti-HCV nAbs. In their work, \timpe found that at \unit{72}{hpi}, $43\pm16$\% of hepatocytes are infected in the presence of nAbs when the free virus transmission route is blocked, relative to that in the absence of nAbs. Our estimated fraction for this data was 91\% (49\%--100\%, 95\% CI) when each data set was weighted appropriately. We investigated a tighter agreement between our simulations and the data presented by \timpe (using a standard deviation of 4\% instead of 16\%) and obtained this way an estimated fraction of 44\% (37\%--53\%, 95\% CI) but found no significant difference in our parameter estimates or the final fractions of cells infected. Our finding that $\sim$1\% of hepatocytes are infected by free virus might appear inconsistent with the observation by \timpe that blocking this mode of transmission reduces the proportion of infection hepatocytes by as much as 57\%, suggesting a larger contribution of free virus transmission to HCV dissemination. However, these findings are not inconsistent when the synergy between these two modes of transmission is considered. 

Herein, we presented visualization of the course of HCV infection wherein the mode by which each hepatocyte became infected is tagged by the ABM. The visualization shows that each new focus is initiated by infection via free virus. Once seeded, a focus grows rapidly and efficiently via direct, cell-to-cell transmission. However, as the focus ages and continues to grow via direct transmission from infected hepatocytes to their immediate neighbours, most hepatocytes which make up the focus find themselves surrounded by other infected hepatocytes: only hepatocytes along the periphery of the focus with yet uninfected neighbours can continue contributing to the focus growth. As Beauchemin explained previously \cite{beauchemin06mixed}, this geometric constraint acts to reduce the effective rate of cell-to-cell transmission as a focus ages. We have observed that most cells in a monolayer have about 6 neighbours on average, and as such we use a hexagonal grid to represent our cells. The natural hexagonal arrangement of confluent hepatocytes in vitro is such that the perimeter ($6n$) to area ($3n(n+1)+1$) ratio of a focus as it ages ($n$) results in an effective cell-to-cell transmission within each focus which is proportional to the inverse of the age of the focus (1/age). Under these circumstances, infection of distal hepatocytes via free virus offers a much needed escape from the smothering effect of the geometric constraint imposed on cell-to-cell transmission. Both modes of transmission work in a coordinated manner: rapid and efficient cell-to-cell transmission is enhanced through a continuously renewed supply of yet uninfected distant hepatocytes made accessible via free virus transmission. While free virus infection accounts for $\sim$1\% of all infections, it accounts for 100\% of new infection foci which are critical to maintaining a high rate of HCV dissemination. 

Our results indeed indicate that the rate of cell-free infection must be extremely low to explain the data. It is known that most cell-free virions are lost without further infection, simply because most virions are not infectious. For HIV, SHIV, RSV, influenza A, HCV, it is common to have 1 infectious virion for every 100 or even every 10,000 or more virions and as such, most cell-free virions will not cause infection. In HCV, this is especially true: to this day, only JFH HCV in Huh7.5(.1) cells can cause in vitro infection and yet HCV continues to be transmitted and to successfully infect humans, yet it barely propagates in vitro. This has also been an issue with the respiratory syncytial virus (RSV) and recently, an experimental in vitro system showed that, at least in that system, most RSV infections occurs directly cell-to-cell, and that cell-free virus are more a consequence/symptom of infected cell death than a significant route of dissemination of the infection to other cells \cite{huong16}. Yet RSV is a highly infectious virus so it is not clear how to reconcile this paradox. Our findings for HCV herein would be consistent with such a scenario.

The experimental data presented herein were all taken from different sources, and as such, the three infections use different viruses/cells. The infection done in \keum used a cell culture derived JFH clone, with modified E2 and p7 protein. The modification of the E2 protein makes the virus more viable (higher concentrations of virus) and the p7 promotes an early stage of viral replication \cite{kim11}. As such, when considering the data from \keum, we did not consider extracting the number of HCV RNA copies per cell or the \trnaoffset from this data since they could be specifically affected by these virus mutations. The three different cell lines, Huh7.5.1, Huh7, and Huh7.5 used by \keum, \sainz, and \timpe, respectively were also considered herein. Huh7.5 and Huh7.5.1 each have mutations reducing the IFN response, making them more susceptible to infection \cite{zhong05}. This should have little effect on our results using \timpe data since this mutation should affect both modes of infection.

The ABM developed herein is not the first mathematical study of HCV cell-to-cell dissemination. \graw developed and evaluated different hypotheses, formulated as mathematical models, for the mechanism of HCV dissemination in vitro. In agreement with past experimental work \cite{timpe08} and with our own findings herein, \graw confirm that free virus transmission alone cannot explain the HCV infection foci distributions observed in vitro. \graw show that claudin-1 and Niemann-Pick C1-like 1 receptors are critical and potentially necessary for cell-to-cell transmission. \graw consider two mathematical models for the rate of cell-to-cell transmission: one proportional to the number of infected hepatocytes in a focus called the birth model, and another proportional to the number of infected hepatocytes along the perimeter of a focus called the boundary model. They find that while both mathematical models could reproduce the experimentally observed distribution of focus sizes, the boundary model better reproduced the frequency of smaller foci, while the birth model performed better for larger focus sizes. This could be because the birth and boundary models considered by \graw both neglect transmission via free virus. Indeed, \graw observe foci consisting of a single hepatocyte as late as \unit{72}{hpi} even in the presence of nAbs. Since new foci can only be formed via free virus colonization, and given that a focus of that size would have to be relatively new and therefore not the result of the initial inoculum, this suggests that perhaps some free virus transmission persists in their experiments despite nAbs. If their boundary model, most similar to our ABM's implementation of cell-to-cell transmission, also included very rare but persistent free virus transmission, it is likely that it would better recapitulate their observed foci size distribution at all focus sizes. Nevertheless, our model is better than the models developed by \graw, as it incorporates both cell-to-cell transmission and transmission via free virus and is able to not only determine that cell-to-cell transmission plays an important role in HCV dissemination but can quantify the relative contributions of each mode of infection. 

While our work establishes the clear dominance of cell-to-cell transmission in efficient HCV dissemination, it is not clear to what extent this finding in vitro would translate in vivo. Some of the geometry of tissue arrangement within the human liver resembles the two-dimensional geometry of the in vitro cell culture. However, a number of host factors, such as interferons, which are part of the innate immune response, could modulate the effectiveness of cell-to-cell transmission in vivo and skew the relative contribution of the two modes. It has been observed that the innate immune response is active within days of HCV infection \cite{wieland05} and Kandathil et al.\ \cite{kandathil13} and Wieland et al.\ \cite{wieland14} both observed multiple focus formations in chronically infected patients, infecting 1\% to 50\% of the liver. As the innate immune response is thought to target cell-free virus, this implies that the innate immune response does not efficiently block all cell-free virus in vivo, but can halt infection to some degree and could affect the relative contribution of the modes of infection. 

If the mode of HCV transmission in vivo resembles that which we establish here in vitro, then the in vitro evaluation of antiviral therapies should enable relevant optimization for compounds which appropriately target both modes of dissemination. If the mode of HCV transmission in vivo differs substantially from that established in vitro, it is likely that in vitro optimization of antiviral targets would be sub-optimal in vivo. In such a case, however, mathematical analyses using our ABM could help discern an antiviral's efficacy with respect to each mode. It could also be used to optimize therapy for any relative contribution of each mode which might exist in vivo. In fact, careful analysis of the differences in antiviral efficacy between the in vitro and in vivo systems could even be used to identify the relative contribution of cell-to-cell versus free-virus dissemination of HCV in vivo.

\cleardoublepage

\subsection*{Acknowledgements}

This work was supported in part by Discovery Grant 355837-2013 (CAAB) from the Natural Sciences and Engineering Research Council of Canada (\url{www.nserc-crsng.gc.ca}), Early Researcher Award ER13-09-040 (CAAB) from the Ministry of Research and Innovation of the Government of Ontario (\url{www.ontario.ca/page/early-researcher-awards}), and by the Interdisciplinary Theoretical and Mathematical Sciences programme (iTHEMS, \url{ithems.riken.jp}) at RIKEN (CAAB). Additional support was provided by AMED (\url{www.amed.go.jp}) through its Japan Program for Infectious Diseases Research and Infrastructure grant numbers 20wm0325007h0001, 20wm0325004s0201, 20wm0325012s0301, 20wm0325015s0301 (SI), its Research Program on Emerging and Re-emerging Infectious Diseases grant numbers 19fk0108156h0001 and 20fk0108140s0801 (SI), its Program for Basic and Clinical Research on Hepatitis grant number 19fk0210036h0502 (SI), and by the Japanese Science and Technology Agency (\url{www.jst.go.jp}) Mirai Program (SI). The funders had no role in study design, data collection and analysis, or decision to publish.

\subsection*{Disclosures}

SI is the Chief Scientific Officer of Science Groove Inc.. Science Groove Inc.\ had no role in the study design, analysis, writing of the manuscript or decision to publish.

\subsection*{Authors contribution}

CAAB, JJF were responsible for study conceptualization and project administration. KB, JJF, SI, CAAB contributed to the development of the methodology. KB, CQ, CAAB contributed to data analysis, software, visualization. CAAB and SI were responsible for funding acquisition. JJF, SI and CAAB were responsible for supervision. KB, JJF, CAAB contributed to the original manuscript draft. All authors contributed to its review and editing.


\cleardoublepage
\section{Methods}

\subsection{Image of Huh7 hexagonal packing from \sainz}

The original image is from the second row, right-most panel of Fig.\ 3 in \cite{sainz09vj}. It is an image of the 2-D monolayer of Huh7 cells seeded in 8-well chamber slides at 80\% confluence and fixed \unit{48}{\hour} post-seeding. The original image was captured by confocal microscopy (630$\times$, Zeiss LSM 510, Germany) and was co-stained with antibodies specific for tight junction protein zona occludens 1 (ZO-1)(Zymed) and Hoechst dye to identify cell nuclei. The digital image file for the figure was obtained directly from the online (web) version of the paper. It was edited using version 0.92.3 of the free and open-source software Inkscape$^\text{TM}$ to copy the image 4 times over, and to colour the outline of the contact neighbourhood of 2 different hepatocytes per image.

\subsection{Parametrization data from \keum}

Data from Fig.\ 1A (black diamond) in \keum, which consists in the mean over three independent experiments of the intracellular HCV RNA per cell over the course of a single-cycle HCV infection (MOI of \unit{6}{\ffu/cell}), were extracted from the electronic version of the paper using version 5 of the software Engauge Digitizer \cite{engaugedigitizer}. The first 3 data points (\unit{3}{\hour}, \unit{6}{\hour}, \unit{9}{\hour}) were discarded because they represent HCV RNA from the inoculum rather than de novo HCV RNA progeny. The last 2 data points (\unit{96}{\hour}, \unit{120}{\hour}) were discarded because intracellular HCV RNA begins to decrease at these points, presumably due to virion progeny export becoming significant. These two phenomena are not captured by Eqn.\ \eqref{intraRNA}, and thus their impact in the data had to be minimized. Our fit of Eqn.\ \eqref{intraRNA} to the remaining data points (\unit{12}{\hour}, \unit{18}{\hour}, \unit{24}{\hour}, \unit{36}{\hour}, \unit{48}{\hour}, \unit{72}{\hour}), determined through nonlinear least-square minimization, yielded best-fit values of $\alpha = \unit{0.097}{\perh}$, $\bar{R} = 1790$, and $\trnaoffset = \unit{10.5}{\hour}$. In the remainder of our work, we only make use of $\alpha$, but not $\bar{R}$ nor $\trnaoffset$ because the latter two are strongly dependent on the experimental system. For example, while this data by \keum suggests a steady-state HCV RNA/cell of $\sim$1800, that from \sainz suggest that number is closer to $\sim$120, as discussed in the text. Furthermore, because Eqn.\ \eqref{intraRNA} does not capture cell infection, \trnaoffset would include a combination of the delay to infect and that to produce progeny, a timing that is further masked by the preexisting intracellular HCV RNA from the inoculum virus in this data set.

\subsection{Parametrization data from \sainz}

Data from \sainz, specifically Fig.\ 3A (open diamond) and 4A (line) for intracellular HCV RNA, and Fig.\ 3B (open diamond) and 4A (bars) for extracellular infectious HCV, were extracted from the electronic version of the paper using version 5 of the software Engauge Digitizer \cite{engaugedigitizer}. According to the figure captions (the paper has no Methods section), the experimental procedures and conditions in their Fig.\ 3 and 4 appear to be identical. Specifically, confluent Huh7 cells treated with 1\% DMSO, replenished every 3 days starting 20 days prior to infection, were infected  with HCV JFH1 at a MOI of \unit{0.01}{FFU/cell}. Data in their Fig.\ 3 are collected over 14 days whereas those in Fig.\ 4 are collected over 62 days. No error bars or replicates are shown in either of their Fig.\ 3 or 4, but the authors state only in their Fig.\ 4 caption that multiple wells were infected and that the data are representative of three independent experiments. The authors do not state whether the ``independent'' experiments in Fig.\ 4 were different wells on the same plate or whether actual distinct experiments were repeated independently. Since no error bars are provided, and since the authors do not explain what they mean by ``representative of'', herein we interpret the data in their Fig.\ 3 and 4 as two distinct and independent repeats of the same experiment. In comparing our ABM against these data, we consider only the early kinetic time points (at 1, 2, 3, 4, 6, 8, 10 days post-infection), after which both intracellular and extracellular HCV reach a steady state value containing no valuable kinetic information. The day 1 and day 2 time points in Fig.\ 3A (open diamond) were moved from \unit{10^0}{FFU/mL} in the original publication to \unit{10^1}{FFU/mL} herein which we believe is the actual limit of detection (LOD). These points were located on the $x$-axis in the original publication, suggesting they were below the LOD. The authors do not provide a methodology or their LOD, but instead refer readers to another of their publication for their methodology \cite{zhong05}, wherein data points below the LOD are also located on the $x$-axis which is now at \unit{10^1}{FFU/mL}.

\subsection{Parametrization data from \timpe}

Parametrization data from Fig.\ 2D (white and grey bars) in \timpe consists in a bar graph reporting the number of JFH1-infected Huh7.5 cells at \unit{72}{hpi} in the presence of anti-HCV nAbs C1, expressed as a percentage of that in the untreated culture. Each bar is given with error bars representing the mean and standard error of the mean (SEM) of 4 replicates which are said to reflect the results of 4 independent experiments. From the digital image of the figure, using version 5 of the software Engauge Digitizer \cite{engaugedigitizer}, we determined these bars to correspond to (mean$\pm$SEM) of (100$\pm$16)\% in the absence of nAbs (white) versus (43$\pm$4)\% in the presence of anti-HCV nAbs (grey). Since the SEM correspond to the standard deviation ($\sigma$) of $N$ measurements, divided by $\sqrt{N}$, i.e.\ SEM $= \sigma/\sqrt{N}$, these measures correspond to a (mean$\pm\sigma$ = mean$\pm$SEM$\sqrt{4}$) of (100$\pm$32)\% and (43$\pm$8)\% in the absence vs presence of nAbs, respectively. Finally, since the number in the presence of nAbs has been divided by the mean in the absence of nAbs, the standard error on the latter must be correctly propagated into that of the former such that
\begin{equation}
\sigma_f = 43\% \left[\left(\frac{32}{100}\right)^2 + \left(\frac{8}{43}\right)^2\right] = 16\%
\label{esigmaf}
\end{equation}
which is how we come to use $(43\pm16)\%$ for this measure, as reported in Table \ref{tdata}. In Figure \ref{freprod}, this quantity is represented graphically as a Gaussian of mean 43\% and standard deviation 16\%. In Eqn.\ \eqref{elikeli} which defines the likelihood of a given parameter set $\vec{p}$, this quantity appears in the numerator as 0.43 and in the denominator as $\sigma_f = 0.16$. We also explore the contribution of this measure to our PPLD estimates by decreasing $\sigma_f$ from 0.16 to 0.04.

\subsection{Reproducing experimental HCV infections with the ABM}

The ABM is implemented in \texttt{python} version 2, and primarily makes use of the \texttt{numpy}, \texttt{random} and \texttt{math} modules. The ABM hexagonal simulation grid contains $(100\times100)$ cells, implemented as a list of cell objects wherein each cell object contains its grid location ($i$,$j$), its state (uninfected, infected by free virus, infected cell-to-cell), a list of its 6 immediate neighbours cell objects, and once infected, its amount of intracellular RNA and the time at which it became infected. In \sainz (specifically, Fig.\ 3 therein), the authors indicate a supernatant infectivity of \unit{10^4}{\ffu/\milli\liter} at time zero corresponding to their reported MOI of \unit{0.01}{\ffu/cell}, suggesting \unit{10^6}{cells/\milli\liter} of supernatant or medium. Furthermore, \sainz (specifically, Fig.\ 2 therein) report a confluent Huh7 density of $\sim$\unit{360,000}{cells/well} on 12-well plates with a culture area of \unit{3.84}{\centi\metre^2/well}. As such, assuming one hepatocyte is ~\unit{20}{\micro \metre} in diameter \cite{lohmann99}, the $(100\times100)$ cells dimension of the ABM grid corresponds to a real-life area of ~\unit{(2\times2)}{\milli\metre} or 1/96 of a 12-well plate well, under a total supernatant volume of \unit{10}{\micro\liter}.

For any given parameter set, \pvec, we first make use of the RNS to initialize the script's random numbers generator instance with a call to \texttt{random.seed(RNS)}. All cells are initialized in the uninfected state, and the extracellular infectious HCV concentration is set to \Vo, one of the ABM parameters to be determined. As explained in \cite{holder11autoimm}, the experimental MOI can be related to the rate of infection by extracellular infectious HCV, $\beta_V$, using
\begin{equation}
\beta_V = \frac{\text{MOI}\cdot c\, \bar{V}}{\Vo \left[1-\me^{-c(\unit{1}{\hour})}\right]}
\label{bveq} \ .
\end{equation}
After \unit{1}{\hour}, the inoculum is rinsed in the ABM following
\begin{equation}
\Vopost = \frinse\, \left[ \Vopre \right] = \frinse\, \left[\Vo\, \me^{-c(\unit{1}{\hour})} \right]\ ,
\label{Vprepost}
\end{equation}
where $\frinse\in[0,1]$ is the effectiveness of the rinse (0 = perfect rinse, 1=no rinse), $c$ is the rate at which cell-free infectious extracellular HCV loses infectivity, and \Vopre and \Vopost are the extracellular infectious HCV concentrations at \unit{1}{\hpi}, immediately prior to and after the rinse, respectively. The rinsing procedure in Eqn.\ \eqref{Vprepost} (two leftmost terms) is repeated every 3 days to simulate the medium replacement as part of replenishing the DMSO treatment described in \sainz. The infection is allowed to progress for 10 days, over which time the intracellular RNA content in each cell summed over all cells, $R(t)$, and the extracellular infectious HCV concentration, $V(t)$, are recorded to be compared to their experimental equivalents shown in Figure \ref{freprod}A,B (or D,E).

Once the infection simulation completes, still using the same $\vec{p}$, the simulation grid is re-initialized (by setting all cells in the uninfected state, and setting the extracellular infectious HCV concentration to \Vo), and the process is repeated a second time to simulate the infection in the presence of anti-HCV nAbs. This infection differs from the one conducted in the absence of nAbs only in the fact that the extracellular infectious HCV concentration, $V(t)$, and rate of HCV release, $p_V(t)$, are set to 0 at $t\ge\unit{8}{\hpi}$, such that $V(t\ge\unit{8}{\hpi})=0$, to simulate the neutralizing effect of nAbs which are introduced at \unit{8}{\hpi} in \timpe. Infection in the presence of nAbs are simulated in the ABM over a period of \unit{72}{\hour} to obtain the number of infected cells at that time. This number divided by that found at the same time in the absence of nAbs, a quantity we denote as \fI, is compared against the ($0.43\pm0.16$) reported for that same quantity by \timpe under equivalent experimental conditions, as shown in Figure \ref{freprod}C (or F).

\begin{figure}
\centering
\resizebox{\textwidth}{!}{\includegraphics{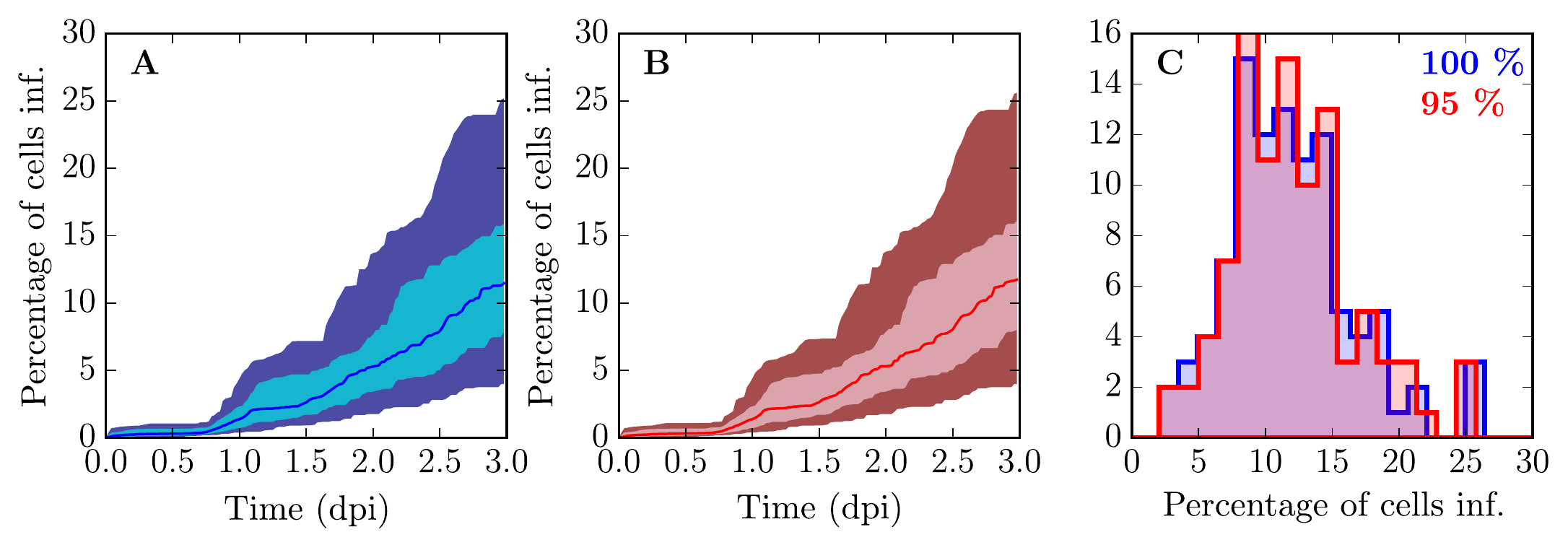}}
\caption{\textbf{Evaluating the impact of perfect (100\%) vs imperfect (95\%) anti-HCV neutralizing antibodies.}
ABM-simulated time courses for the \textbf{(A)} percentage of all HCV-infected cells where we assumed perfect (100\%) nAbs efficacy. \textbf{(B)} Same as (A) but we assumed an imperfect (95\%) efficacy. The solid line (blue or red) indicate the median, and the pale and darker bands (blue or red) indicate the 68\% and 95\% percentiles of the $y$-value at each $x$-value (time point) for 100 ABM simulations performed using parameter sets pulled at random, with replacement from the PPLD. \textbf{(C)} Histogram of the percentage of infected cells at \unit{3}{dpi} of the ABM simulations, where we assumed perfect (100\%, blue) or imperfect (95\%, red) nAbs efficacy. The effect of imperfect nAbs efficacy is not significant given the inter-simulation variability due to stochasticity. %
}\label{fnabs}
\end{figure}

The procedure described above assumes perfect nAbs efficacy (100\%). \timpe demonstrated that their nAbs' efficacy is no less than 95\%. For completeness, we also evaluated whether a maximally imperfect neutralizing efficacy of 95\% would significantly affect the contribution of free virus infection obtained using a perfect efficacy of 100\%. To do so, 100 parameter sets were pulled at random, with replacement from the PPLD, and used to simulate $\unit{3}{\dday}$ infections in the presence of perfect (100\%) or maximally imperfect (95\%) nAbs efficacy. Figure \ref{fnabs} shows the resulting time course simulations and shows that the effect of the nAbs efficacy is insignificant in the face of the ABM inter-experimental variability.

\subsection{Estimation of posterior parameter likelihood distributions via MCMC}

The parameters estimated are (\trnaoffset, $c$, $\beta_C$, $V(0)$, MOI, \frinse, RNS). The likelihood of a particular parameter set is defined, as in our previous work \cite{beauchemin17}, as $\mathcal{L}(\vec{p}) \propto \exp[-\ssr(\vec{p})/(2\sigma) ]$ where
\ben
\frac{\ssr(\vec{p})}{\sigma} = \sum_{vi=1}^{13} \left[\frac{\log_{10}\left(\frac{V_{vi}^\text{model}(\vec{p})}{V_{vi}^\text{data}}\right)}{\sigma_V}\right]^2 + \sum_{ri=1}^{13} \left[\frac{\log_{10}\left(\frac{R_{ri}^\text{model}(\vec{p})}{R_{ri}^\text{data}}\right)}{\sigma_R}\right]^2 + \left[\frac{\fI(\vec{p})-0.43}{\sigma_f}\right]^2 \ .
\label{elikeli}
\een
The first two terms correspond to the infection kinetic data from \sainz, while the last term corresponds to the ($43\pm16$)\% measure extracted from \timpe (see Table \ref{tdata}). Specifically, $V_{vi}^\text{model}(\vec{p})$ and $V_{vi}^\text{data}$ correspond to the ABM-simulated and experimentally measured extracellular infectious HCV concentrations of the ${vi}^\mathrm{th}$ time point measurement, $R_{ri}^\text{model}(\vec{p})$ and $R_{ri}^\text{data}$ correspond to the ABM-simulated and experimentally measured intracellular HCV RNA concentration of the ${ri}^\mathrm{th}$ time point measurement, and \fI is the ratio of ABM-predicted HCV-infected hepatocytes at \unit{72}{\hpi} in the presence of anti-HCV nAbs divided by that in their absence compared to the ratio of 0.43 reported by \timpe. Each of the three terms is divided by its standard deviation, namely $\sigma_V$, $\sigma_R$, and $\sigma_f$, in order to give it the appropriate weight in the likelihood measure. Since \sainz do not provide error bars for their data, we performed nonlinear least-square minimization to determine a best-fit $\vec{p}$, and computed the standard deviation of the residuals between the best-fit curve and the data, namely $\log_{10}\left(V_{vi}^\text{model}(\vec{p})/V_{vi}^\text{data}\right)$ and $\log_{10}\left(R_{ri}^\text{model}(\vec{p})/R_{ri}^\text{data}\right)$. Through this procedure, we found $\sigma_V = 0.43$ and $\sigma_R = 0.47$, which correspond to an estimated experimental variability of about half an order of magnitude (3-fold or half a $\log_{10}$). These estimates are consistent with the spread of $\sim$3-fold ($10^{\sigma}$) between the pair of experiments (stars and circles) performed by \sainz, shown in Figure \ref{freprod}(A,B). We set $\sigma_f = 0.16$ as per Eqn.\ \eqref{esigmaf}.

The Markov chain Monte Carlo (MCMC) method implemented in version 2.1.0 of the \texttt{emcee} python module \cite{emcee} was used to estimate the parameters' posterior likelihood distribution (PPLDs) based on the likelihood Eqn.\ \eqref{elikeli}. The \texttt{emcee} MCMC implementation is such that at each new step for each chain, given a current position $\vec{p}_\text{current}$, a new parameter set $\vec{p}_\text{new}$ is accepted and added to the chain with a probability proportional to $\mathcal{L}(\vec{p})$, and if rejected $\vec{p}_\text{current}$ is added instead to the chain \cite{emcee}. Our MCMC run consists in 168 chains (or walkers), each 3000 steps in length, out of which we discard the first 1000 steps as burn-in, for a total of 336,000 accepted parameter sets. The burn-in is necessary to remove any dependence of the PPLD on the initial positions chosen for the 168 chains. Though irrelevant, the initial position of each chain was chosen randomly from a normal or log-normal distribution about the best-fit $\vec{p}$, determined through a preliminary nonlinear least-square fit, as in \cite{beauchemin17}.

For all parameters, we assume positive uniform priors $\in[0,\infty)$, with the following additional constraints on parameters. The offset in the replication of intracellular HCV RNA by newly infected hepatocytes, \trnaoffset, was restricted to $>\,$\unit{1}{h}. The rate of HCV loss of infectivity $c\in\unit{[0.01,0.3]}{\perh}$, the widest possible biologically reasonable range, where \unit{0.01}{\perh} is the typical degradation rate of vRNA (which is much more stable than infectivity), and \unit{0.3}{\perh} is, to our knowledge, the largest reported infectivity decay rate for HCV in vitro \cite{song10}. The cell-to-cell transmission rate was restricted to $\beta_C\in[0,\unit{100}{\perh}]$, since $\beta_C > \unit{100}{\perh}$ corresponds to the instantaneous infection of any cell with a cell-to-cell transmission-capable neighbour, with no further change in infection kinetics for larger values. The initial extracellular infectious HCV concentration, \Vo, was restricted to $>\,$\unit{10^1}{FFU/mL}, i.e.\ the limit of detection. The MOI was restricted to $>\,$\unit{10^{-5}}{FFU/cell}, i.e.\ 1000-fold below the experimentally-reported MOI of \unit{0.01}{FFU/cell}. The efficacy of the inoculum rinse was restricted to $\frinse\in(10^{-6},1]$, since $\frinse < 10^{-6}$ yields identical results, and $\frinse = 1$ corresponds to the least efficient rinse possible, i.e.\ no rinsing at all. Based on \texttt{python}'s \texttt{random} module, RNS must be a long integer $\in(0,8\times16^{15})$. 

\subsection{ABM-predicted time course of HCV infection}

The intracellular HCV RNA (A and D), extracellular virus (B and E) from Figure \ref{freprod} and the different fractions of cells infected in Figure \ref{ffractiming} are representative of $500$ parameter sets. These sets were pulled at random, with replacement from the PPLD, and simulated using the ABM for $\unit{10}{d}$. The red line (or blue line) represents the median, and the pale and darker bands (grey or cyan) indicate the 68\% and 95\% percentiles of the $y$-value at each $x$-value (time point) for the ABM simulations performed.

\subsection{Comparison of ABM-predicted foci size distribution}

Additional data extracted from Fig.\ 1E,F (black bars) in \timpe using Engauge Digitizer \cite{engaugedigitizer}, were used to provide validation of the behaviour predicted by the parametrized ABM. In their Fig.\ 1E (black bars), \timpe report counting frequencies of 0.48 small ($<$10 cells), 0.26 medium (10--50 cells), and 0.26 large ($>$50 cells) foci out of a total of 50 foci, observed at \unit{72}{\hpi} in the absence of nAbs. In their Fig.\ 1F (black bars) \timpe report counting frequencies of 0.04 small, 0.26 medium, and 0.70 large foci out of a total of 50 foci, observed \unit{72}{\hpi} in the presence of agar. In both experiments, Huh7.5 cells seeded at high density were infected with HCV-JFH1 at a MOI of 0.01 and no error bars were provided on the bar graphs, presumably because all 50 foci considered were from a single replicate, though this is not specified in their paper. The actual experimental measurements taken by \timpe were not frequencies, but counts (24 small, 13 medium and 13 large foci out of 50 foci counted in the absence of nAbs and 2 small, 13 medium and 35 large in the presence of agar).

 In reproducing this experiment with the parametrized ABM, 1000 parameter sets were pulled at random, with replacement from the PPLD, and used to simulate a \unit{72}{\hour} infection in the absence of nAbs, or presence of nAbs, producing a spatial grid like that shown in Figure \ref{fsshot}. For each grid we counted the number of foci, and the number of cells in each foci. Any grouping of cells in direct contact (i.e.\ within one another's 6-cell neighbourhood) were considered part of the same focus, without discriminating between a single large focus or one resulting from two or more smaller merged foci, analogous to the experimental procedure followed by \timpe. In \timpe, the foci size distribution is affected by cell division (doubling time $\approx$ \unit{34}{h} reported therein) and should be appropriately accounted for when counting our fixed number of cells. Hence, we counted each cell in a foci once if it had been infected for less than \unit{34}{h}, twice if it had been infected between \unit{34}{h} and \unit{68}{h}, and 4 times if infected longer than \unit{68}{h}. The corresponding frequency of small, medium and large foci for each grid is represented in Figure \ref{ftimpeclouds}.


\subsection{Placing a larger emphasis on the \timpe data}

\begin{figure}
\resizebox{0.95\textwidth}{!}{\includegraphics{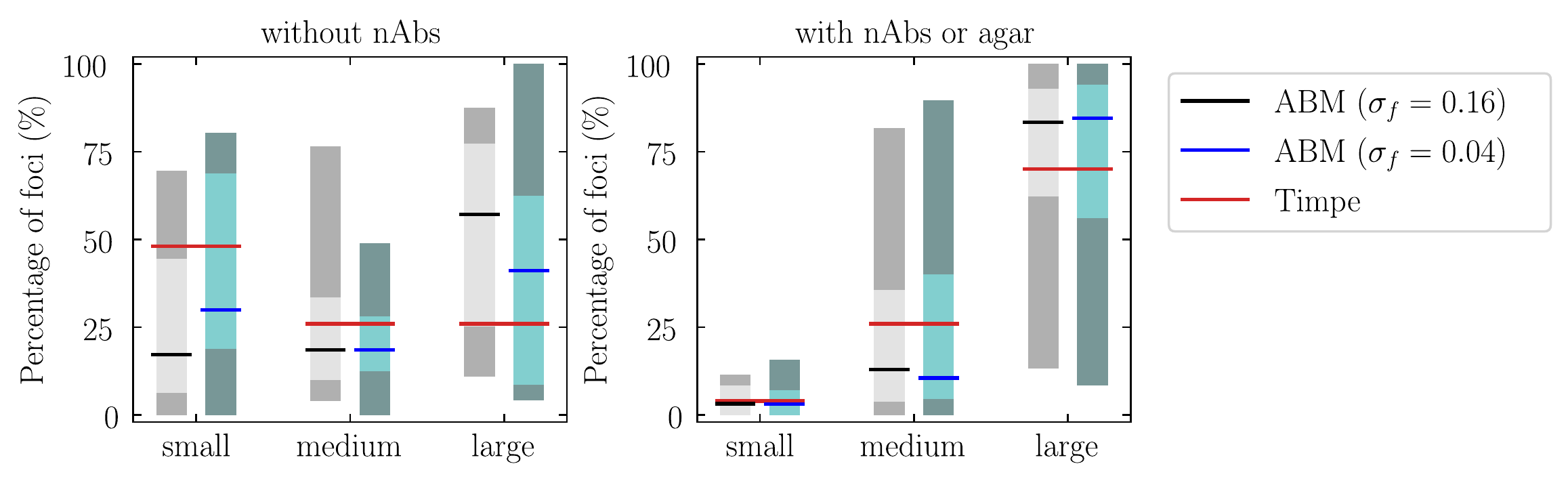}}
\caption{%
\textbf{Quantitative comparison of ABM-predicted foci size distributions between our two MCMC simulations.} The ABM-simulated percentage of small ($<10$ cells), medium (10--50 cells), and large ($>50$ cells) foci at \unit{72}{hpi} for 1000 simulations, shown as percentile bars (median, 68\% CI, 98\% CI = bar and bands) from MCMC simulation with an appropriate weight on the \timpe data (black bar and grey bands, $\sigma_f=0.16$) and from MCMC simulation with a larger weight on the \timpe data (blue bar and bands, $\sigma_f = 0.04$). These are compared against that observed by \timpe (red bar). Results are shown for HCV infections in the absence (left) and presence (right) of free virus dissemination, which was blocked in our ABM by adding nAbs, whereas \timpe used agar. See Methods for details on how the ABM-predictions were performed.%
}
\label{fpredict_04}
\end{figure}

\begin{figure}
\resizebox{\textwidth}{!}{\includegraphics{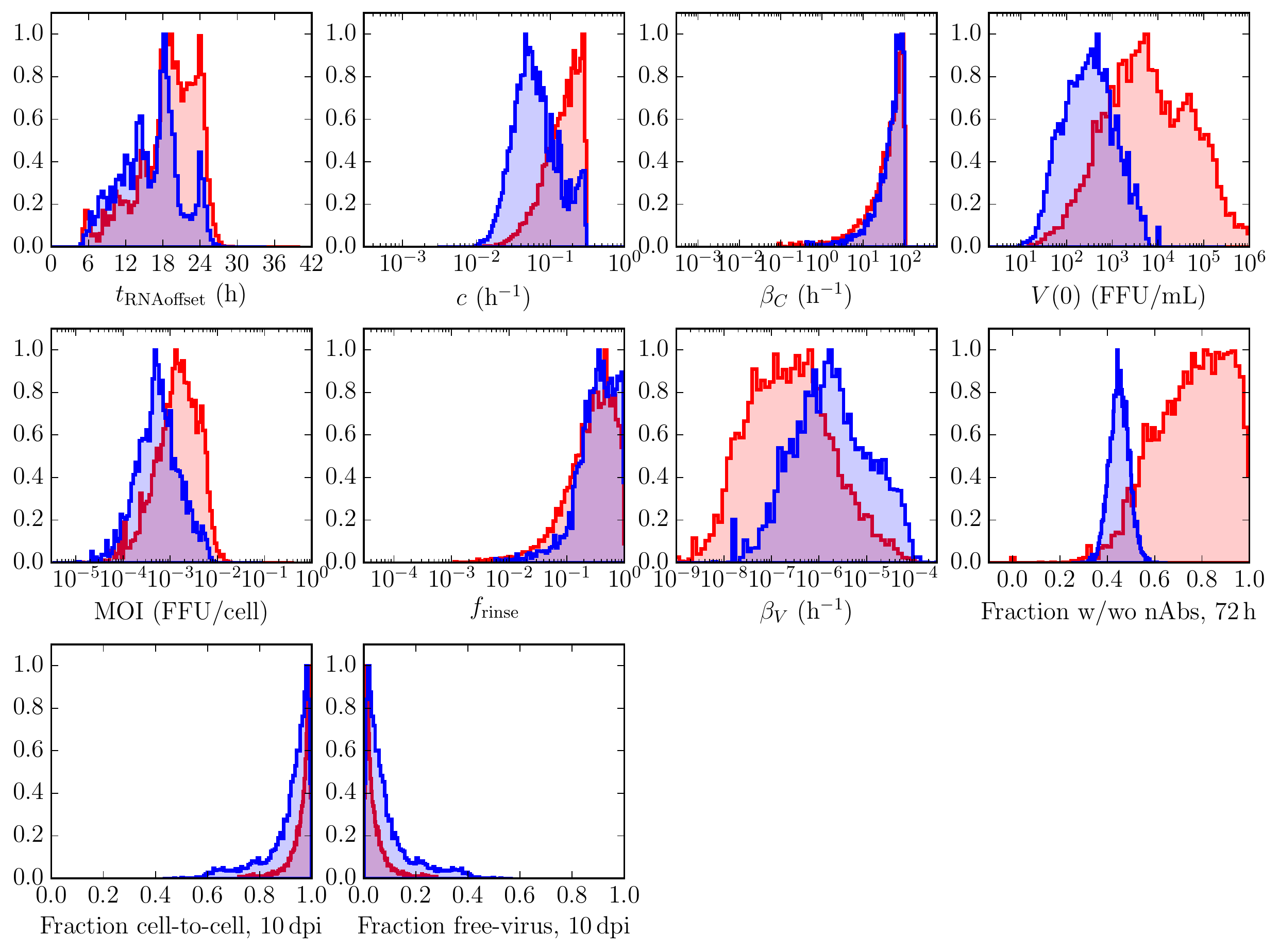}}
\caption{%
\textbf{Comparison of the ABM parameters between our two MCMC simulations.}
Simulated predictions of parameters from appropriately weighted MCMC (red, $43\pm16$\%) and the artificially enforced MCMC (blue, $43\pm4$\%).  
}
\label{param_compare}
\end{figure}

To determine how the \timpe measure of ($43\pm16$\%) affects the results, the MCMC PPLD estimation was repeated with 16$\times$ more weight (using a standard deviation of 4\% instead of 16\%) on this measure. With the exception of this weight factor, the process was identical to that performed with the physically correct standard deviation for this measure (16\%). Our distribution of foci size is similar to our previous results (Figure \ref{fpredict_04}). From our simulated parameters, apart from the fraction of cell-to-cell infection, the only noticeable difference is observed of the clearance rate and the initial concentration of virus (Figure \ref{param_compare}). This is not unexpected since these two parameters can be determined from features of the extracellular virus concentration. Since the relative weight on the extracellular virus has been reduced, this would allow for a bit more freedom with any parameter which is determined by this data. The overall findings for the fraction of cell infected by cell-to-cell yields similar mode values with a wider distribution (98\% [70\%--100\%, 95\% CI] compared to 99\% [84\%--100\%, 95\% CI]) (Table \ref{tmoretimpe}).

\begin{table}
\begin{center}
\caption{MCMC-estimated PPLDs with normal and larger weight placed on the \timpe data.$^a$}
\label{tmoretimpe}
\begin{tabular}{llll}
 & $\sigma_f=0.04^b$ & $\sigma_f=0.16^b$ & fold-change$^c$ \\ [0.2ex]
\hline
\multicolumn{3}{c}{--- PPLDs estimated via MCMC ---} \\
Intracellular HCV RNA replication delay, \trnaoffset (\hour) &
	$18$ $[7.3, 25]$ & $19$ $[9.2,26]$ & $0.79$ $[0.32,2.1]$ \\
Rate of loss of HCV infectivity, $c$ (\perh) &
	$0.049$ $[0.023,0.30]$ & $0.27$ $[0.049,0.30]$ & $0.30$ $[0.087,2.6]$ \\
Rate of infection via cell-to-cell, $\beta_C$ (\perh) &
	$85$ $[6.5,100]$ & $85$ $[3.6,100]$ & $0.99$ $[0.057,31]$ \\
Initial virus concentration, \Vo (\Vu) &
	$10^{2.7\ [1.5, 3.6]}$ & $10^{3.7\ [2.0, 5.5]}$ & $0.073$ $[0.00039,5.7]$ \\
Multiplicity of infection, MOI (\ffu/cell) &
	$10^{-3.3 [-4.0, -2.2]}$ & $10^{-2.9\ [-3.8, -2.0]}$ & $0.36$ $[0.024,7.9]$ \\
Effectiveness of the rinse, \frinse &
	$0.37$ $[0.062,1.0]$ & $0.47$ $[0.036,1.0]$ & $1.0$ $[0.086,23]$ \\
\hline
\multicolumn{3}{c}{--- Computed from MCMC PPLDs (see Methods) ---} \\
Rate of infection via free virus, $\beta_V$ (\perh) &
	$10^{-5.8\ [-7.1, -4.2]}$ & $10^{-6.5\ [-8.2, -4.9]}$ & $9.6$ $[0.038,1600]$ \\
Percent infected cells w/wo nAbs at \unit{72}{\hpi}, \fI (\%) &
	$44$ $[37, 53]$ & $91$ $[49,100]$ & $0.53$ $[0.40,0.98]$ \\
Percent of infections via cell-to-cell at \unit{10}{dpi} (\%) &
	$98$ $[70, 100]$ & $99$ $[84,100]$ & $0.99$ $[0.65,1.2]$ \\
Percent of infections via free virus at \unit{10}{dpi} (\%) &
	$2.1$ $[0.25, 30]$ & $1.2$ $[0.038,16]$ & $2.3$ $[0.11,37]$ \\
\hline \\[-1.7em]
\end{tabular}
\end{center}
$^a$ Normal weight on \timpe data is indicated by $\sigma_f=0.16$ and larger weight by $\sigma_f=0.04$. \\
$^b$ Mode and 95\% CI determined from the PPLDs.\\
$^c$ Power of 10 of the mode and 95\% CI determined from the distribution of $\log_{10}$(fold-change) between 1,680,000 pairwise comparisons, wherein each pair is formed by drawing one value from the PPLD with a larger ($\sigma_f=0.04$) and normal ($\sigma_f=0.16$) weight placed on the \timpe data, sampled at random with replacement.
\end{table}

\cleardoublepage
\bibliographystyle{abbrvurl}
\bibliography{hcv-distal-local}

\end{document}